\documentclass[twocolumn]{aastex631}

\usepackage{gensymb}
\usepackage{amsmath}
\usepackage{xspace}
\usepackage{xcolor}
\usepackage[caption=false]{subfig}

\newcommand{\edot}{\ensuremath{\dot{E}}\xspace}

\newcommand{\tcE}{\ensuremath{\tau_{c}(E)}\xspace}

\newcommand{\db}{\ensuremath{D_{B}}\xspace}

\def\amin{\ifmmode^{\prime}\else$^{\prime}$\fi}
\def\asec{\ifmmode^{\prime\prime}\else$^{\prime\prime}$\fi}

\newcommand{\hess}{HESS~J1826$-$130}

\newcommand{\hawc}{HAWC~J1826$-$128}
\newcommand{\psr}{PSR~J1826$-$1256\xspace}

\newcommand{\nustar}{\textit{NuSTAR}}

\newcommand{\xmm}{{\it XMM}}
\newcommand{\xmmnewton}{{\it XMM-Newton}}
\newcommand{\chandra}{{\it Chandra}}
\newcommand{\fermi}{{\it Fermi-}LAT\xspace}

\def\asec{\ifmmode^{\prime\prime}\else$^{\prime\prime}$\fi}

\def\simgt{\lower.5ex\hbox{$\; \buildrel > \over \sim \;$}}
\def\simlt{\lower.5ex\hbox{$\; \buildrel < \over \sim \;$}}

\shorttitle{The Eel PWN}
\shortauthors{Burgess et al.}
\graphicspath{{./}{figures/}}

\begin{document}

\title{The Eel Pulsar Wind Nebula: a PeVatron-Candidate Origin for \hawc\ and \hess}

\author{Daniel~A.~Burgess}\affiliation{Columbia Astrophysics Laboratory, 550 West 120th Street, New York, NY 10027, USA}
\author{Kaya~Mori}\affiliation{Columbia Astrophysics Laboratory, 550 West 120th Street, New York, NY 10027, USA}
\author{Joseph~D.~Gelfand}\affiliation{NYU Abu Dhabi, PO Box 129188, Abu Dhabi, United Arab Emirates}
\author{Charles~J.~Hailey}\affiliation{Columbia Astrophysics Laboratory, 550 West 120th Street, New York, NY 10027, USA}
\author{Yarone~M.~Tokayer}\affiliation{Columbia Astrophysics Laboratory, 550 West 120th Street, New York, NY 10027, USA}
\author{Jooyun~Woo}\affiliation{Columbia Astrophysics Laboratory, 550 West 120th Street, New York, NY 10027, USA}
\author{Hongjun~An}\affiliation{Chungbuk National University, Chungdae-ro 1, Seowon-Gu, Cheongju, Chungbuk, 28644 South Korea}
\author{Kelly~Malone}\affiliation{Physics Division, Los Alamos National Laboratory, Los Alamos, NM, USA}
\author{Stephen~P.~Reynolds}\affiliation{Physics Department, NC State University, Raleigh, NC 27695, USA}
\author{Samar~Safi-Harb}\affiliation{Department of Physics and Astronomy, University of Manitoba, Winnipeg, MB R3T 2N2, Canada}
\author{Tea~Temim}\affiliation{Department of Astrophysical Sciences, Princeton University, Princeton, NJ 08544, USA}

\begin{abstract}

\hawc\ is one of the brightest Galactic TeV $\gamma$-ray sources detected by the High Altitude Water Cherenkov (HAWC) Observatory, with photon energies extending up to nearly $\sim$100~TeV. This HAWC source spatially coincides with the H.E.S.S. TeV source \hess\ and the ``Eel" pulsar wind nebula (PWN), which is associated with the GeV pulsar \psr. In the X-ray band, \chandra\ and \xmmnewton\ revealed that the Eel PWN is composed of both a compact nebula ($\sim$15\asec) and diffuse X-ray emission ($\sim$6\arcmin$\times$2\arcmin) extending away from the pulsar. Our \nustar\ observation detected hard X-ray emission from the compact PWN up to $\sim$20 keV and evidence of the synchrotron burn-off effect. In addition to the spatial coincidence between \hess\ and the diffuse X-ray PWN, our multi-wavelength spectral energy distribution (SED) analysis using X-ray and $\gamma$-ray data establishes a leptonic origin of the TeV emission associated with the Eel PWN. Furthermore, our evolutionary PWN SED model suggests (1) a low PWN  B-field of $\sim$1 $\mu$G, (2) a significantly younger pulsar age ($t \sim5.7$ kyr) than the characteristic age ($\tau= 14.4$ kyr) and (3) a maximum electron energy of $E_{\rm max} = 2$~PeV. The low B-field as well as the putative supersonic motion of the pulsar may account for the asymmetric morphology of the diffuse X-ray emission. Our results suggest that the Eel PWN may be a leptonic PeVatron particle accelerator powered by the $\sim$6--kyr--old pulsar \psr\ with a spin-down power of $3.6 \times 10^{36}$ erg s$^{-1}$. 

\end{abstract}


\section{Introduction}
\label{sec:intro}

The origin of extremely energetic cosmic rays (CRs) above $\sim$1 PeV is a long-standing problem in high energy astrophysics. Since the directional information of each CR is washed out by interstellar magnetic fields, the energetic CR acceleration in our Galaxy can be probed indirectly through their candidate accelerators, which are frequently associated with TeV $\gamma$-ray sources. 

The High-Altitude Water Cherenkov (HAWC) experiment has opened a new window to the high energy $\gamma$-ray sky. Equipped with 300 water Cherenkov detectors (WCDs), HAWC directly detects air shower particles produced by very high energy (VHE, $E\ge100$~GeV) up through ultra high energy (UHE, $E\ge100$~TeV) $\gamma$-rays in the upper atmosphere \citep{Smith2015}.
The HAWC observatory is distinct from imaging air Cherenkov telescopes such as
H.E.S.S., VERITAS and MAGIC due to its sensitivity at energies above $\sim$10 TeV, a result of employing WCDs and collecting data from $\gamma$-ray sources continuously under all weather conditions. Alongside the LHAASO (Large High Altitude Air Shower Observatory) experiment, HAWC explores astrophysical sources in the highest energy $\gamma$-ray band beyond $\sim$100 TeV. A handful of HAWC sources detected at and beyond $\sim$100~TeV are of particular interest as they may be produced by one of the most extreme classes of cosmic particle accelerator -- a {\it Galactic PeVatron} \citep{Abeysekara2020}. Primary Galactic PeVatron candidates that may accelerate electrons and protons to PeV energies are pulsar wind nebulae \citep[][]{Arons2012}, $\gamma$-ray superbubbles \citep[e.g. the Cygnus cocoon;][]{Hona2021} and the supermassive black hole Sgr A* \citep{2016Natur.531..476H}.

While the number of Galactic PeVatron candidates detected by HAWC is growing, it remains uncertain whether leptonic or hadronic accelerators are primarily powering their UHE TeV emission. 
We now have strong indications that CR electrons and positrons are generated by leptonic processes in pulsar wind nebulae (PWNe), are able to exit these PWNe, and may diffuse to great distances \citep{Cholis2018}, in part due to the recent discovery of TeV $\gamma$-ray halos around the Geminga and Monogem pulsars \citep{Abeysekara2017, Linden2017}. The existence of $\gamma$-ray halos and successful leptonic model fits to UHE HAWC sources \citep[e.g.][]{Sudoh2021} both support a leptonic PWN hypothesis for these UHE HAWC sources and, by association, for Galactic PeVatrons and the origin of CRs. 

PWNe are composed of highly relativistic particle winds that are injected by energetic rotation-powered pulsars into the surrounding SNR ejecta or interstellar medium. 
High energy observations have detected synchrotron and inverse Compton scattering (ICS) emission from numerous PWNe in the X-ray and TeV bands respectively, suggesting that non-thermal electrons are accelerated to GeV--PeV energies within PWNe. Given the complex emission mechanisms and morphology of TeV sources, the origins of TeV emission are best explored through multi-wavelength observations from the radio, X-ray, GeV, and up to TeV energy bands. For example, PWNe are often found in the vicinity of supernova remnants (SNRs) and molecular clouds, and therefore can be difficult to spatially resolve with $\gamma$-ray observations alone. 

Over the past two decades, X-ray and TeV $\gamma$-ray observations have bolstered our understanding of PWN astrophysics and electron acceleration in the Galaxy. In a typical PWN, synchrotron radiation dominates from radio to keV/MeV photon energies, while the TeV emission is produced by ICS of accelerated electrons off CMB photons or local IR/optical/UV radiation fields from nearby dust, molecular clouds and star clusters \citep[e.g.][]{Slane2017}. 
As of today, more than 70 X-ray emitting PWNe have been discovered and nearly half of them have been detected in the TeV band  \citep{Kargaltsev2013}. \chandra\ observations of PWNe have revealed detailed X-ray features such as tori, jets, and bow-shocks \citep{Kargaltsev2015}. On the other hand, \nustar, operating in its broad bandwidth of 3 to 79~keV, has detected hard X-ray synchrotron emission from some of the highest energy electron populations in our Galaxy ($E_e > 100$~TeV), including PWNe and X-ray filaments in the Galactic Center \citep{Zhang2014, Mori2015, Reynolds2017}.

\begin{figure*}
    \centering
    \subfloat
    {\includegraphics[width=1.03\columnwidth]{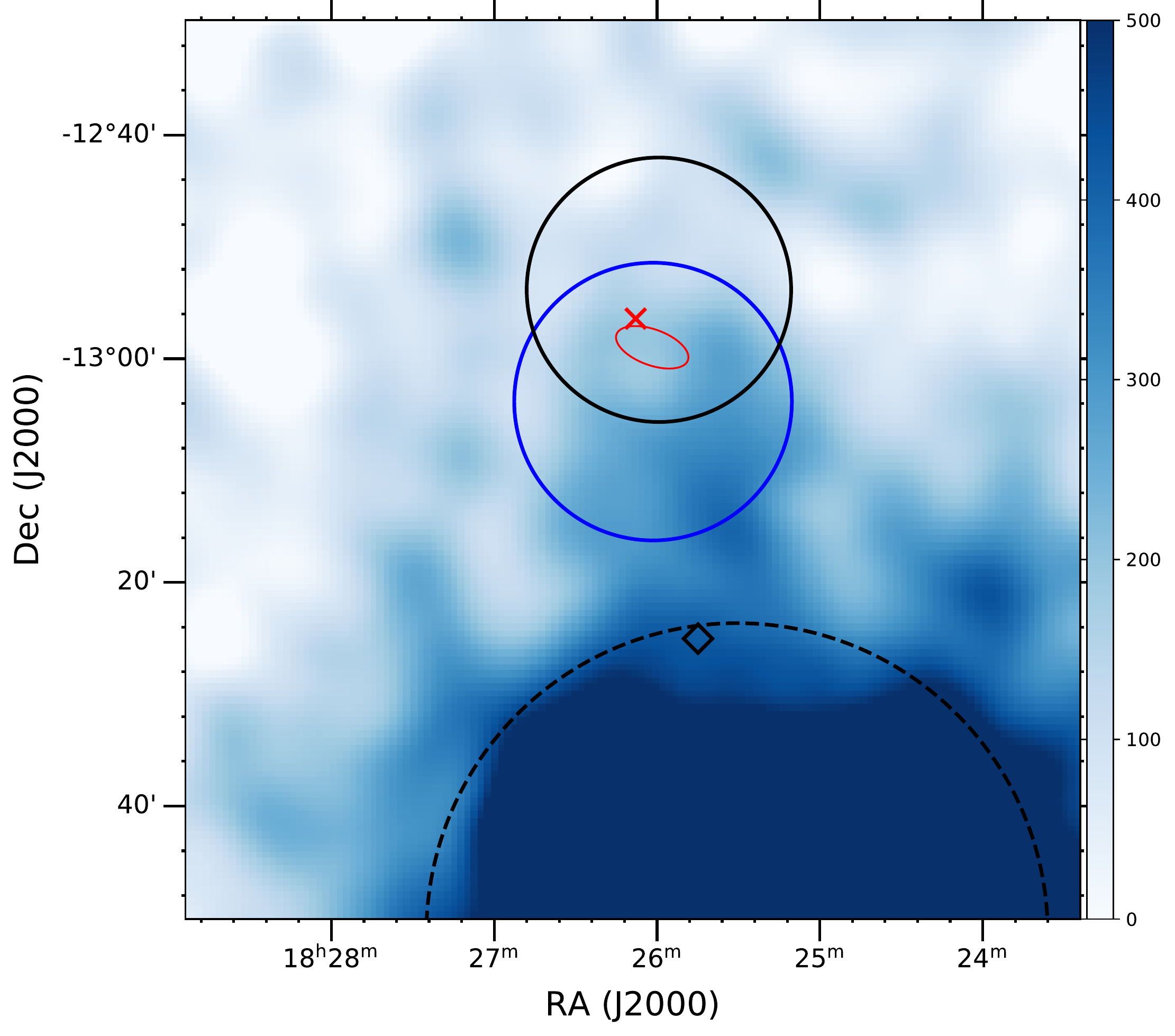}}
    \qquad
    \subfloat
    {{\includegraphics[width=1.0\columnwidth]{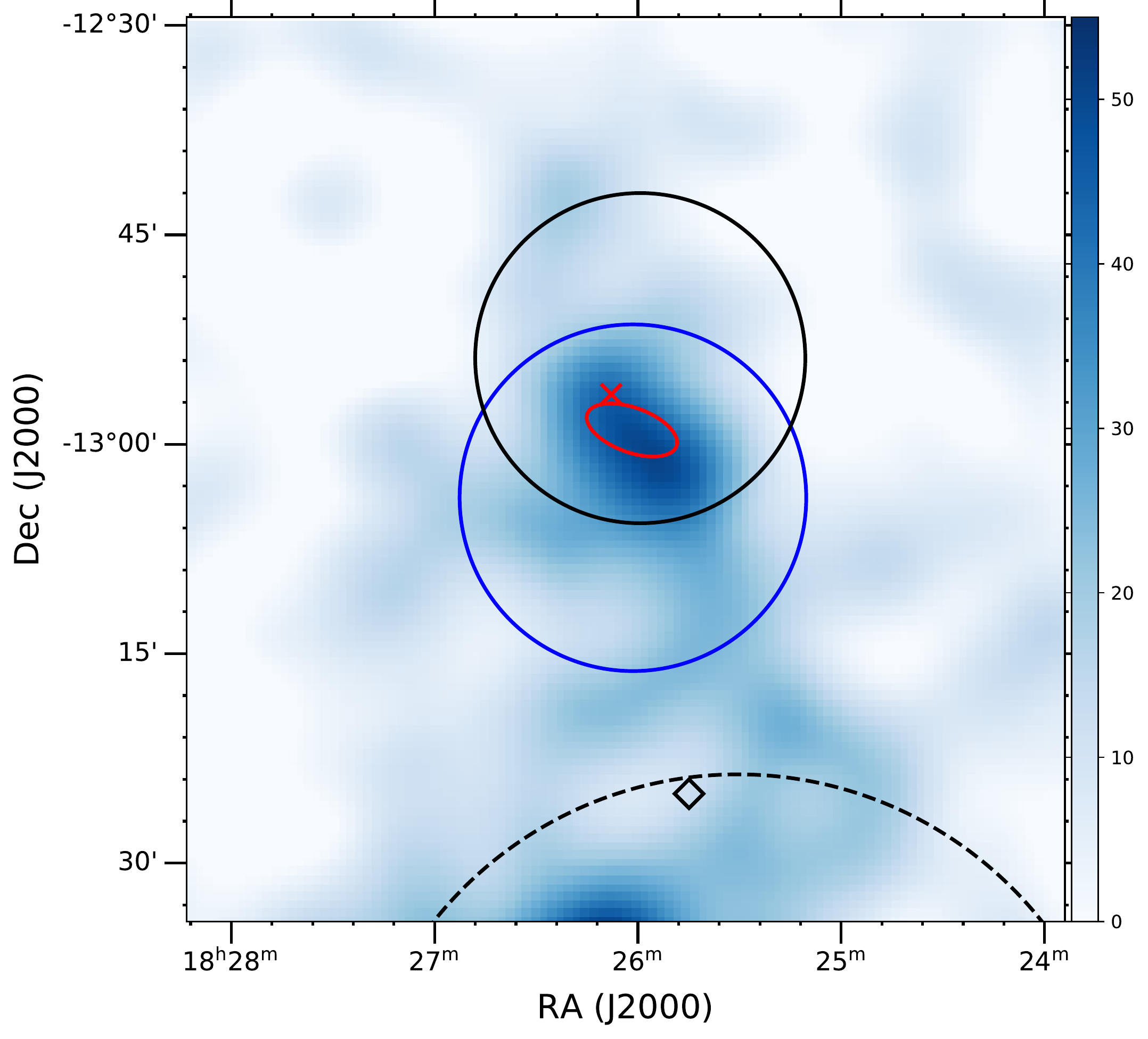}}}
    \caption{H.E.S.S. 1--10 TeV (\textit{left panel}) and $>$ 10 TeV (\textit{right panel}) excess maps \citep[published in][]{Abdalla2019} of the region surrounding \psr\ (location marked with a red cross). Both maps are smoothed to 68\% point spread function (PSF) containment, r$=$0.05\degree. The \hawc\ extension is shown with a solid black circle, the HAWC J1825$-$134 point source is marked with a black diamond, and the HAWC J1825$-$138 extension to the south is shown with a dashed black circle. The blue circle marks the \hess\ extension, showing that the H.E.S.S and HAWC source are spatially coincident with \psr. The extended and bright TeV emission to the South is HESS J1825$-$137, the likely counterpart to HAWC J1825$-$138. The $E>10$ TeV excess map clearly shows TeV emission spatially associated with the diffuse Eel PWN (marked by the red ellipse).}
    \label{fig:region}
\end{figure*}

The highly extended UHE source eHWC J1825$-$134 belongs to a small group of nine $\gamma$-ray sources detected above 56 TeV, registered in the \textit{energetic} HAWC (eHWC) source catalog \citep{Abeysekara2020}. This eHWC detection corresponds with LHAASO J1825$-$1326, a new UHE $\gamma$-ray detection over 1$\degree$ in extension with a photon E$_{max}$ of 0.42 $\pm$ 0.16 PeV \citep{Cao2021}. A more recent HAWC analysis was able to resolve eHWC J1825$-$134 into three separate high energy TeV sources, one of which is \hawc\ \citep{Albert2021}. 
This HAWC source is spatially coincident with the known TeV source \hess, which may imply that \hawc\ is a separate detection of this TeV accelerator, providing a higher energy component to the original H.E.S.S. source. \cite{Albert2021} found that the \hawc\ spectrum is hard in TeV $\gamma$-rays, reaching photon energies of nearly 100 TeV. \hess\ also features a relatively hard spectrum \citep{Abdalla2020} and the similarly hard spectral indices of both sources ($\Gamma < 2$) further supports their association. 
At larger transverse distances, \hawc\ and \hess\ are adjacent to the extended and softer TeV emission from HAWC J1825$-$138 in the south (the likely counterpart to HESS J1825$-$137) and HAWC J1825$-$134, a newly-discovered $\gamma$-ray source (detected above $\sim$200 TeV) likely powered by hadronic particle acceleration \citep[see Figure \ref{fig:region} and][]{Albert2021}.

\begin{figure*}%
    \centering
    \subfloat {\includegraphics[width=0.48\textwidth]{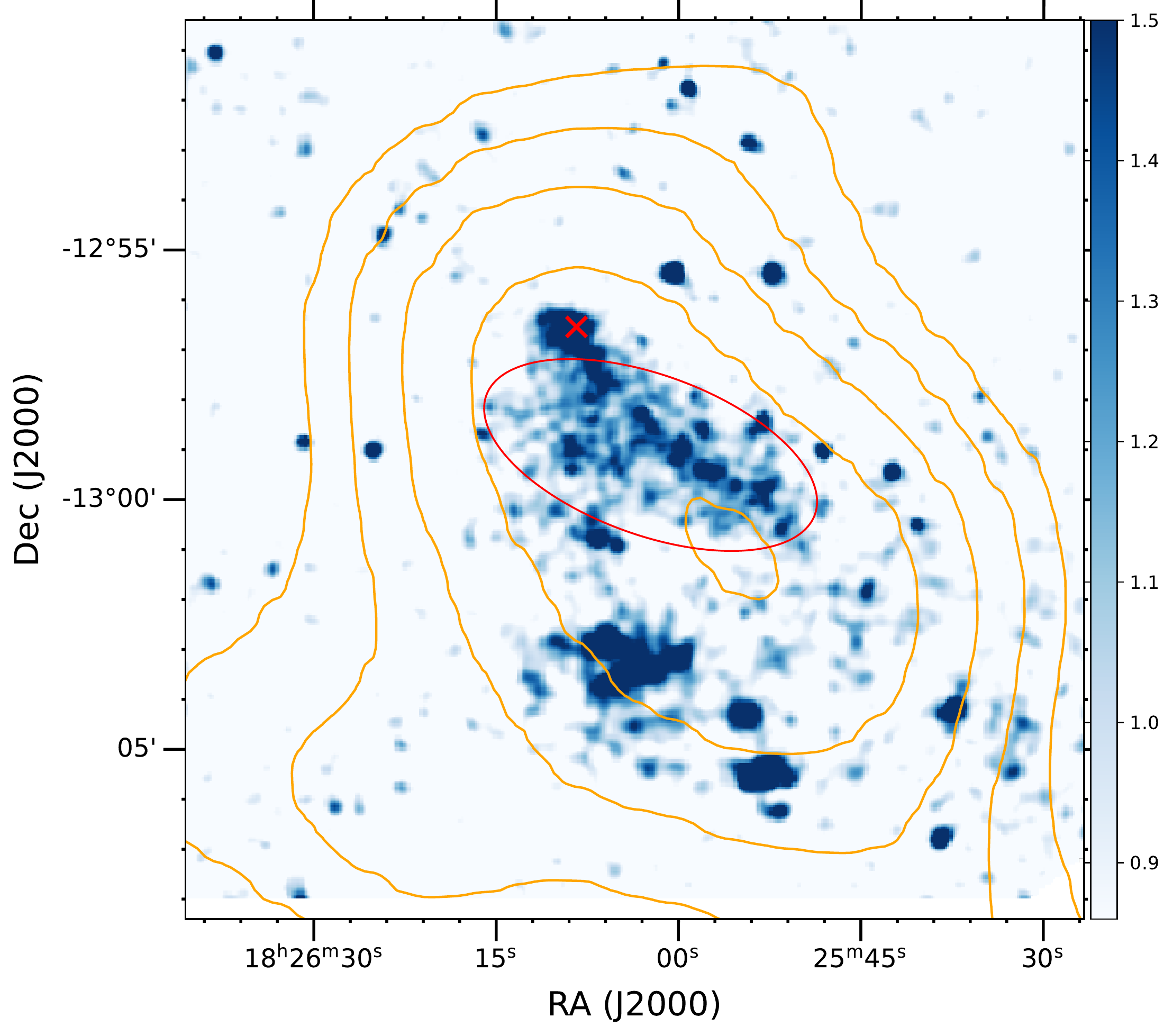}}%
    \qquad
    \subfloat {{\includegraphics[width=0.48\textwidth]{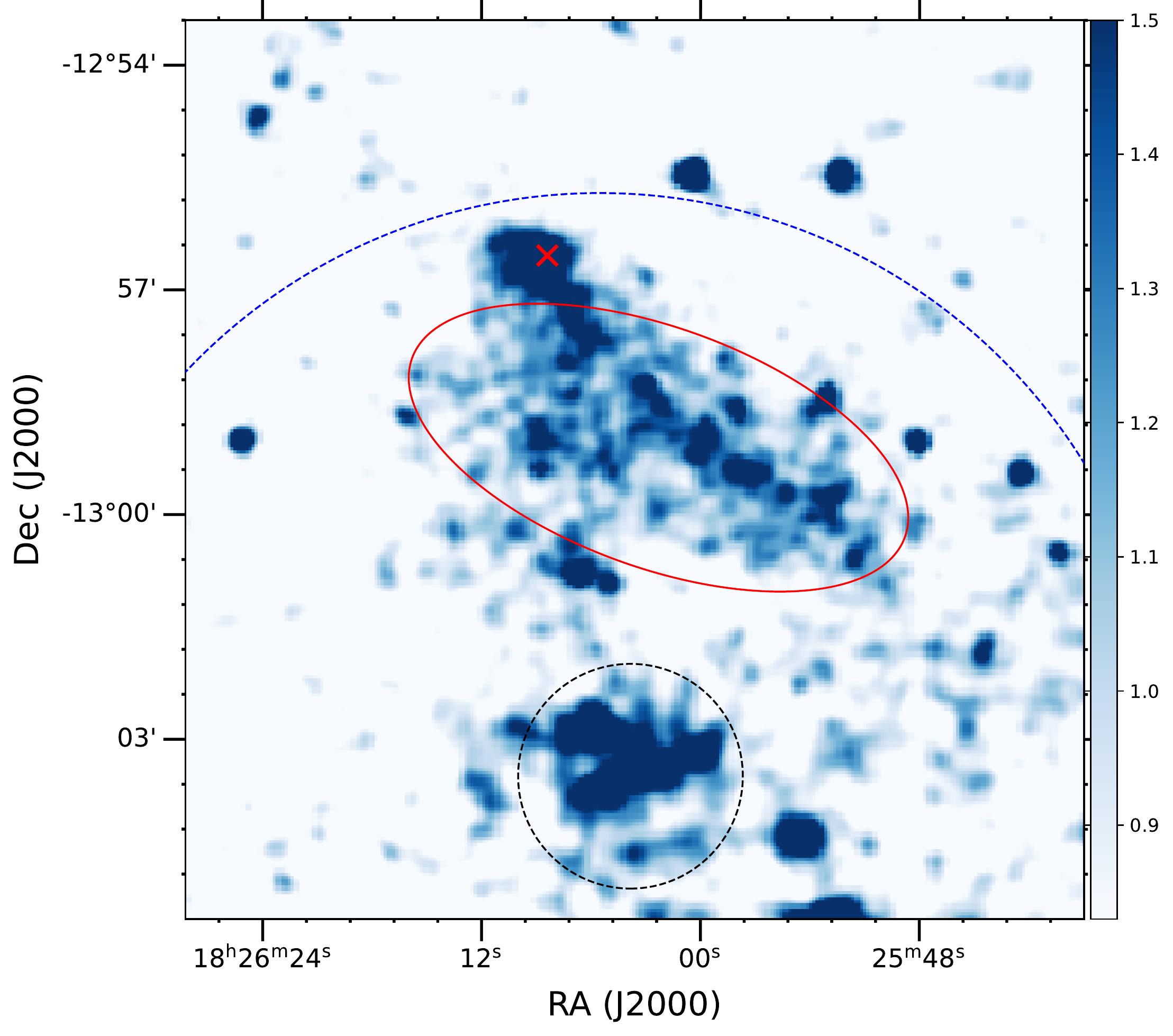}}}%
    \caption{\textit{Both panels:} \xmmnewton\ 0.5--10 keV images after merging MOS1 and MOS2 data, background subtraction and exposure correction. The images are adaptively smoothed by a Gaussian kernel with $\sigma = 3$ pixels, and are scaled to a weighted count rate per pixel via \texttt{eimageget} defaults. \textit{Left panel:} H.E.S.S. source contours (min. 3$\sigma$ to 5$\sigma$ significance, from 20 to 50 cts pixel$^{-1}$) above 10 TeV are displayed in orange, showing $\gamma$-ray association with the diffuse PWN (see \S\ref{sec:extended}). \psr's location is marked with a red cross. The red ellipse shows the region used for diffuse PWN spectral analysis, as well as all SED fitting (\S\ref{sec:extended}, \S\ref{sec:sed}). \textit{Right panel:} the same \xmmnewton\ MOS image zoomed in around the Eel PWN (with an 11\arcmin $\times$11\arcmin\ FOV). The dashed blue circle shows the radio shell of candidate SNR G18.45$-$0.42 \citep{Anderson2017}, and open cluster Bica 3 is marked with a dashed black circle.}%
    \label{fig:xmm}%
\end{figure*}

A GeV pulsar (\psr) of high spin-down luminosity \edot\ and its diffuse PWN are roughly coincident with \hawc\ (pulsar 5\amin.4 offset from $\gamma$-ray centroid) and \hess\ (pulsar 4\amin.6 offset from the $\gamma$-ray centroid). 
\psr\ is one of the brightest GeV $\gamma$-ray pulsars in the Fermi 4-year catalog 3FGL \citep{Acero2015}, and has been categorized as ``Vela-like'' \citep{Karpova2019}. Located at R.A.(J2000) = 18$^h$ 26$^m$ 08.5$^s$ decl.(J2000) = --12$\degree$ 56$'$ 33.36$''$, \psr\ has a characteristic age of 14.4 kyr, spin-down luminosity \edot\ of $3.6 \times 10^{36}$ erg s$^{-1}$, and surface magnetic field $B = 3.7 \times 10^{12}$ G \citep{Kargaltsev2017}. 
\xmmnewton\ and \chandra\ X-ray observations have characterized \psr\ and the surrounding PWN \citep[known as the Eel PWN;][]{Roberts2007} in the X-ray band below 10 keV. Previous analyses of these \xmmnewton\ and \chandra\ observations and the various X-ray and TeV $\gamma$-ray properties of the Eel PWN include \cite{Karpova2019} and \cite{Duvidovich2019}. In the X-ray band, the Eel PWN is characterised by a tail-like extended emission over 6\amin\ (see Figure \ref{fig:xmm}) and a more compact nebula within $r\sim15$\asec\ from the pulsar \citep{Karpova2019}. Hereafter we refer to them as ``diffuse PWN'' and ``compact PWN'', respectively. Additionally, \cite{Karpova2019} used the relation between distance and interstellar reddening to derive a new \psr\ distance estimate of 3.5 kpc, which we adopt throughout this paper for all transverse distance calculations (e.g. 1\amin\ $\sim$ 1D$_{3.5kpc}$ pc).

Motivated by a potential association between the HAWC and H.E.S.S. sources and the Eel PWN, we performed a \nustar\ observation of the PWN in order to obtain broad-band X-ray data and combine it with the existing \xmm\ data, as well as HAWC/H.E.S.S. TeV data. Our primary goal is to explore the hypothesis that the Eel PWN may be the PeVatron candidate \hawc/\hess\ using multi-wavelength spectral energy distribution (SED) fitting and an analysis of the PWN characteristics and morphology in the X-ray band. 
We begin by reviewing previous observations of the HAWC and H.E.S.S. sources (\S\ref{sec:tev obs}), the status of \fermi\ detection of the pulsar and \hess\ (\S\ref{sec:fermi}), the X-ray observations performed by \nustar\ and \xmmnewton\ (\S\ref{sec:xray obs}), and a 90 cm (330 MHz) radio image of the Eel PWN region (\S\ref{sec:radio}). In \S\ref{sec:imaging} we perform analysis on the \nustar\ X-ray images, and in \S\ref{sec:xrayspec} we perform spectral analyses across the three X-ray telescopes. In \S\ref{sec:sed} we investigate scenarios that may describe the observed emission from \hawc\ and \hess\ through SED fitting, and in \S\ref{sec:disc} we discuss what is powering \hawc\ and \hess, the implications of our fitting results, our conclusions about the Eel PWN morphology, and areas that require further study.

\section{TeV $\gamma$-ray Observations}
\label{sec:tev obs}

The High Altitude Water Cherenkov (HAWC) observatory uses 300 water Cherenkov detectors (WCDs) to detect TeV $\gamma$-rays by proxy of air shower particles. \hawc\ is located at R.A.(J2000) = 18$^h$ 26$^m$ 00.0$^s$ decl.(J2000) = $-$12$\degree$ 51$'$ 36.0$''$, with a Gaussian width of 0.2$\degree$. This source can be fit (energy range of 2$^{+3}_{-1}$ to 84$^{+30}_{-25}$ TeV) with an exponential cut-off power-law model with a spectral index $\Gamma$ = $1.2^{+0.4}_{-0.4}(stat.)^{+0.4}_{-0.5}(syst.)$ and a cut-off energy of $24 ^{+10}_{-7}(stat.)^{+15}_{-7}(syst.)$ TeV \citep{Albert2021}.

The High Energy Stereoscopic System (H.E.S.S.) is comprised of five imaging atmospheric Cherenkov telescopes. The VHE H.E.S.S data of \hess\ we use were obtained with the phase I array configuration, which features an angular resolution better than 0.1$\degree$ \citep{Aharonian2006}. Observations of the region containing \hess\ occurred between 2004 and 2015, resulting in a corrected live-time of 206 hours \citep{Anguner2017}. The centroid of \hess\ is located at R.A.(J2000) = 18$^h$ 26$^m$ 02.6$^s$ decl.(J2000) = --13$\degree$ 00$'$ 55.0$''$, with a source extension of 0.21$\degree$ \citep{Abdalla2020}. This source is best fit (energy range of 0.5$^{+0.1}_{-0.1}$ to 43$^{+13}_{-11}$ TeV) with an exponential cut-off power-law model, which gives a spectral index $\Gamma$ $= 1.78 \pm 0.10$ and a cut-off energy of approximately 15 TeV, \citep{Abdalla2020}. 

The centroids of \hawc\ and \hess\ are offset by approximately 10\arcmin, and both source extensions overlap with the diffuse Eel PWN (see Figure \ref{fig:region}). Because of this spatial coincidence, along with the spectral consistency between the two sources within uncertainties, we proceed under the assumption that these are two separate detections of the same $\gamma$-ray source, as suggested by \cite{Albert2021}.


\begin{figure*}[ht!]%
    \centering
    \subfloat {\includegraphics[width=0.48\textwidth]{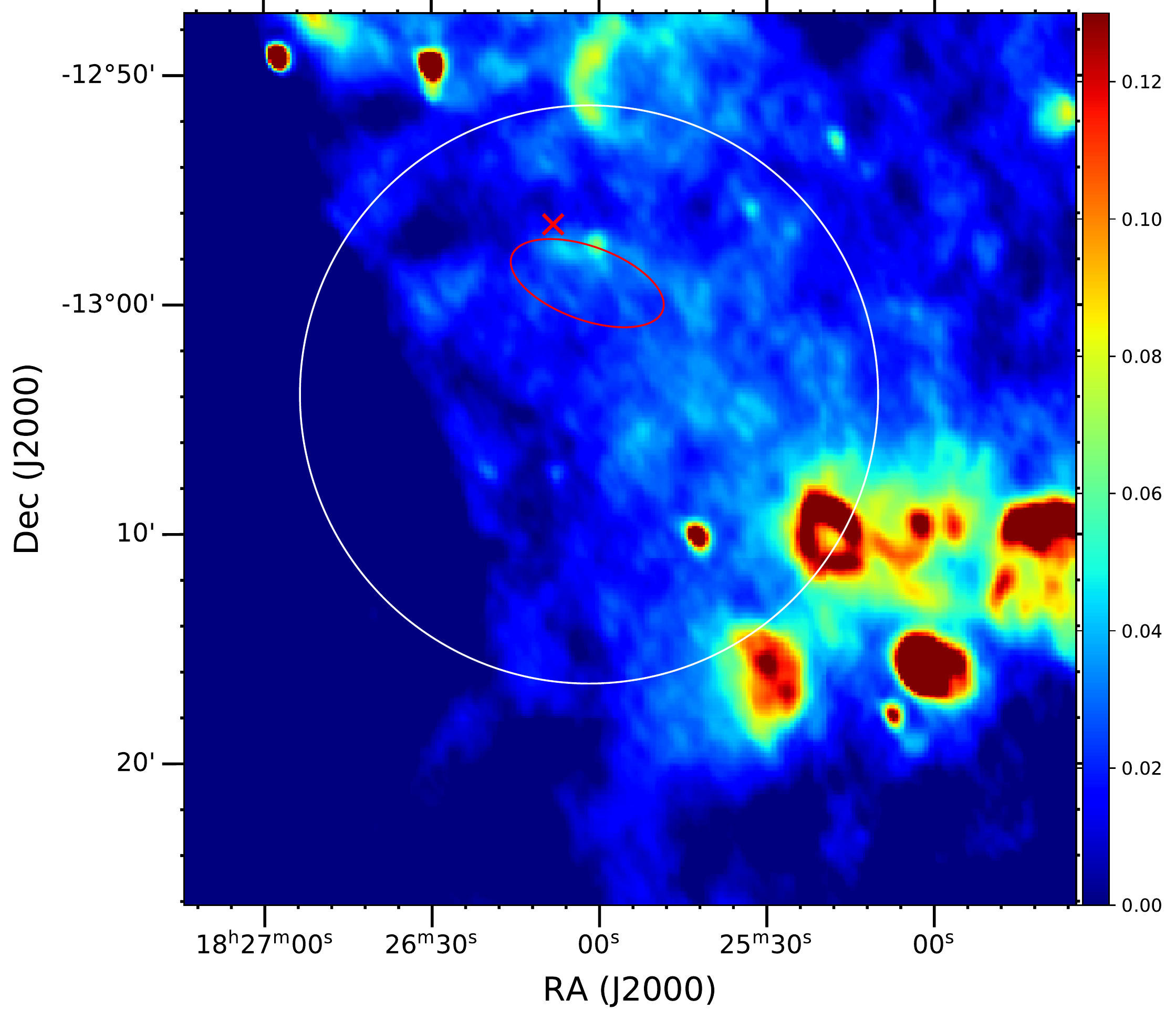}}%
    \qquad
    \subfloat {\includegraphics[width=0.48\textwidth]{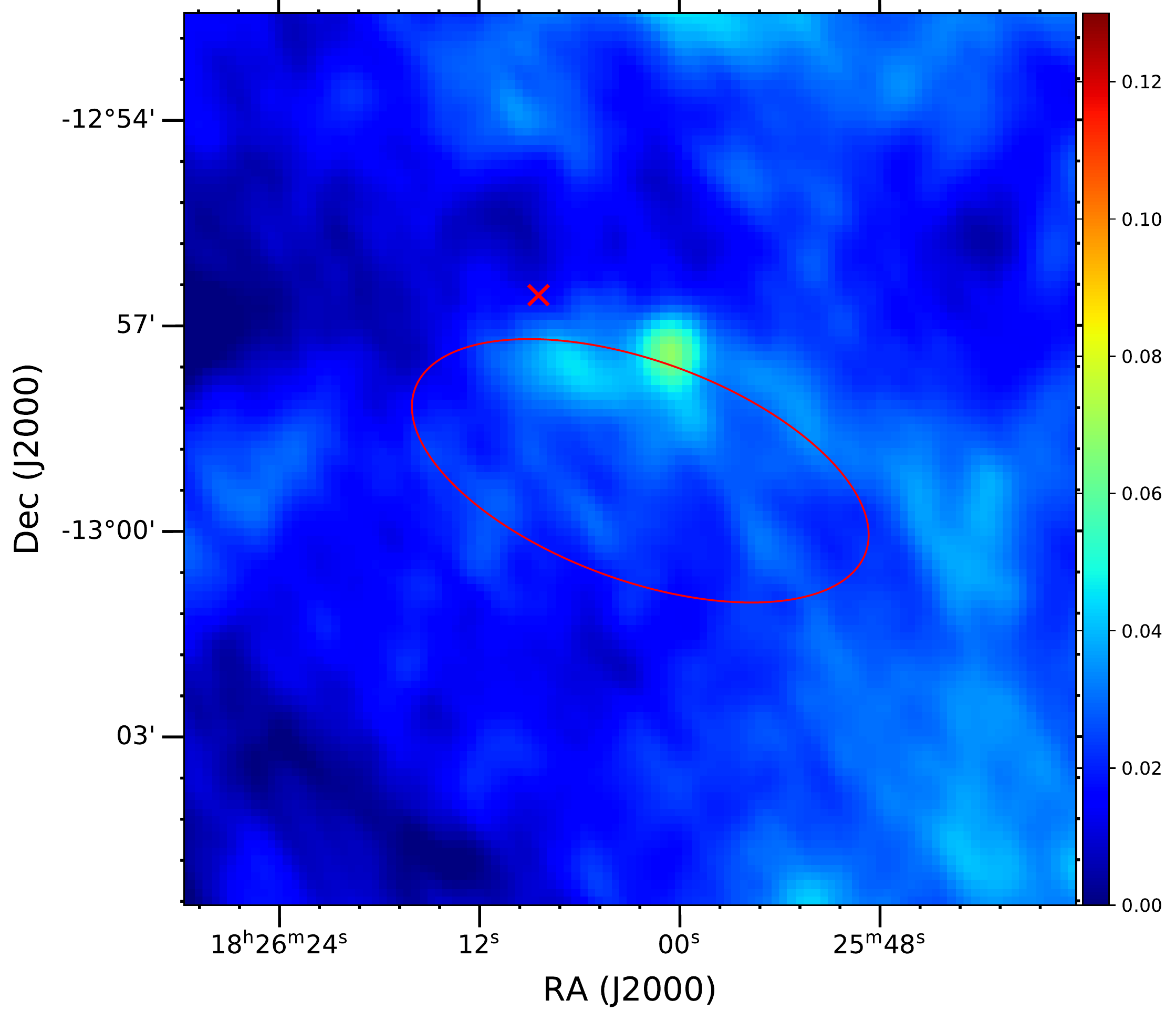}}%
    \caption{\textit{Both panels:} a 90 cm (330 MHz) VLA radio image of the region surrounding the Eel PWN in pixel units of Jy/beam. \textit{Right panel:} a zoomed in perspective. The location of \psr\ is marked with a red cross, and the diffuse Eel PWN source region from Figure \ref{fig:xmm} is marked by a red ellipse. The \hess\ source region from Figure \ref{fig:region} is shown with a white circle. A slight enhancement of 90 cm radio emission is visible near the pulsar and is partially coincident with the diffuse PWN. The H II region GAL 018.14$-$00.29 and several unidentified NRAO VLA Sky Survey (NVSS) sources are visible in the southwest of the wide field image.}%
    \label{fig:radio}%
\end{figure*}

\section{\fermi\ Observations and Data Analysis}
\label{sec:fermi}

The pulsar associated with the Eel PWN (\psr) is registered in both the 3FHL ~\citep{Ajello2017} and 4FGL \citep{Abdollahi2020} catalogs, as 3FHL/4FGL J1826.1$-$1256. To investigate whether \fermi\ detected any counts from the location of \hess\ that could not be attributed to the pulsar, we analyzed 13 years of {\it Fermi} Large Area Telescope (LAT) data collected between August 4, 2008, and August 3, 2021 ({\it Fermi} mission elapsed time 239557417s $-$ 649724483s). We used the latest Pass 8 version of the data with ``Source" class (evclass=128) ``FRONT+BACK" type (evtype=3) events in the energy range from 100 MeV to 500 GeV. We used the instrument response function \texttt{P8R3\_SOURCE\_V3} and the filter expression \texttt{(DATA\_QUAL>0)\&\&(LAT\_CONFIG==1)}. To limit the $\gamma$-ray contamination from Earth’s limb, the maximum zenith angle was set to 90$\degree$. The region of interest (ROI) is a square area of 10$\degree\times$10$\degree$ centered at the position of 4FGL J1826.1$-$1256. All point-like and extended sources within a square area of 30$\degree\times$30$\degree$ from the center of the ROI listed in LAT 10-year Source Catalog (4FGL-DR2) were included in the model. The Galactic diffuse emission and extragalactic isotropic emission were modeled using \texttt{gll\_iem\_v07.fits} and \texttt{iso\_P8R3\_SOURCE\_V3\_v1.txt}, respectively\footnote{\url{https://fermi.gsfc.nasa.gov/ssc/data/access/lat/BackgroundModels.html}}. We undertook the binned likelihood analysis with \texttt{Fermipy} ~\citep{Wood2017} version 1.0.1. 

\psr\ is a young, energetic pulsar with emission detected above 20 GeV ~\citep{Ajello2017}. Therefore, we generated test statistics (TS) maps above 30 GeV, 40 GeV, and 50 GeV to detect any emission that could not be attributed to the pulsar. All TS maps showed less than 1$\sigma$ TS values in the region containing \psr\ and \hess. This non-detection of \hess\ is in agreement with ~\cite{Principe2020}, who used off-pulse data from 1 GeV to 1 TeV and did not find evidence of residuals. We derived GeV flux upper limits for 4FGL J1826.1$-$1256 using the \texttt{sed()} method of \texttt{Fermipy}. In this method, the normalization is fit in each energy bin using a power-law model with the index = 2. The 95\% confidence level upper limits were evaluated from the profile likelihood, which yielded a 30--52 GeV upper limit of $8.49 \times 10^{-13}$ erg cm$^{-2}$ s$^{-1}$ and a 52--91 GeV upper limit of $1.51 \times 10^{-12}$ erg cm$^{-2}$ s$^{-1}$. We use these flux upper limits in our spectral energy distribution (SED) fits in \S\ref{sec:sed}-\S\ref{sec:evolutionary}.

\section{X-ray Observations and Data Reduction}
\label{sec:xray obs}

\psr\ and the surrounding Eel pulsar wind nebula have been observed by X-ray telescopes including \nustar\ and \xmmnewton.


\begin{figure*}%
    \centering
    \subfloat {\includegraphics[width=0.47\textwidth]{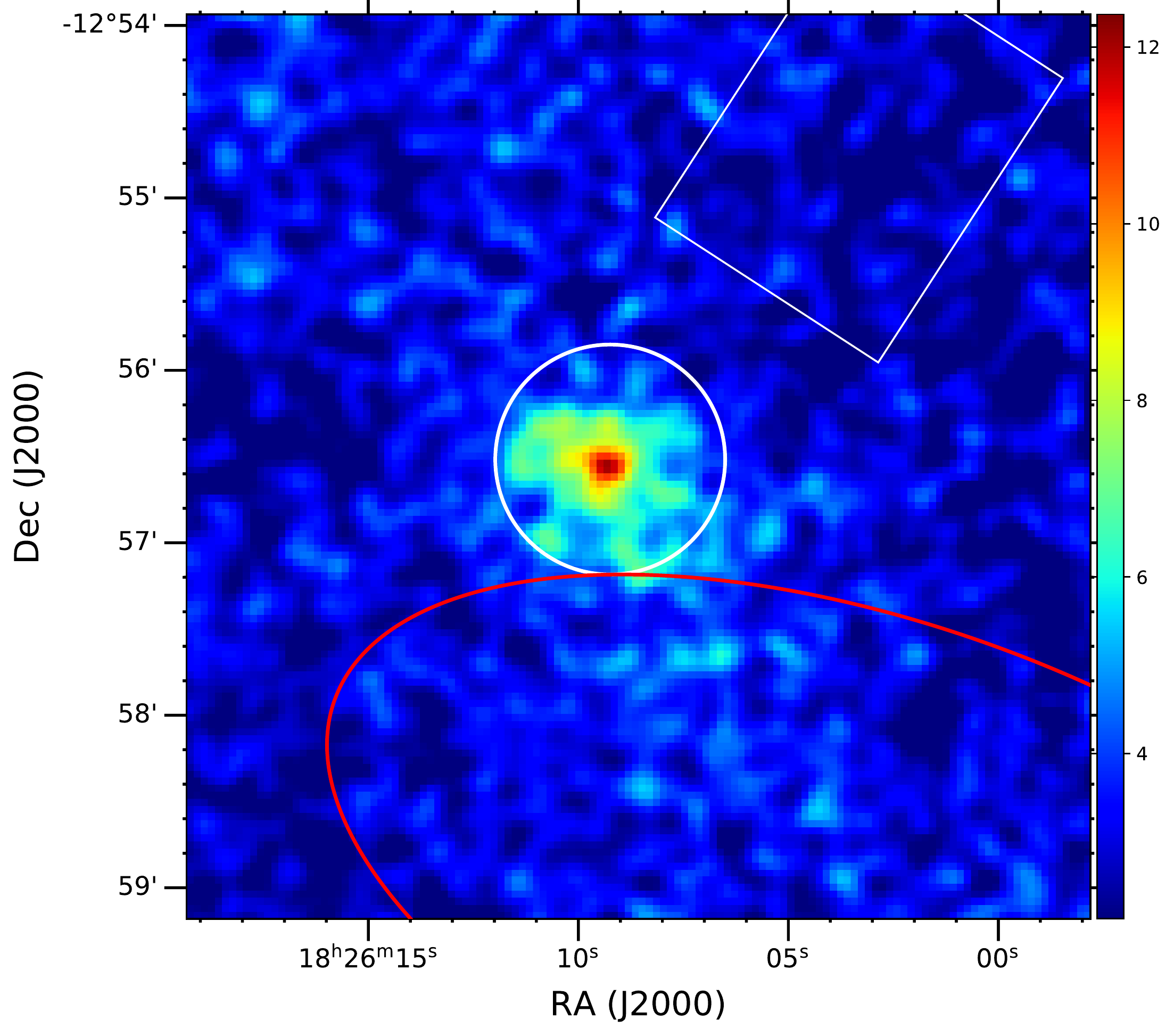}}%
    \qquad
    \subfloat {{\includegraphics[width=0.47\textwidth]{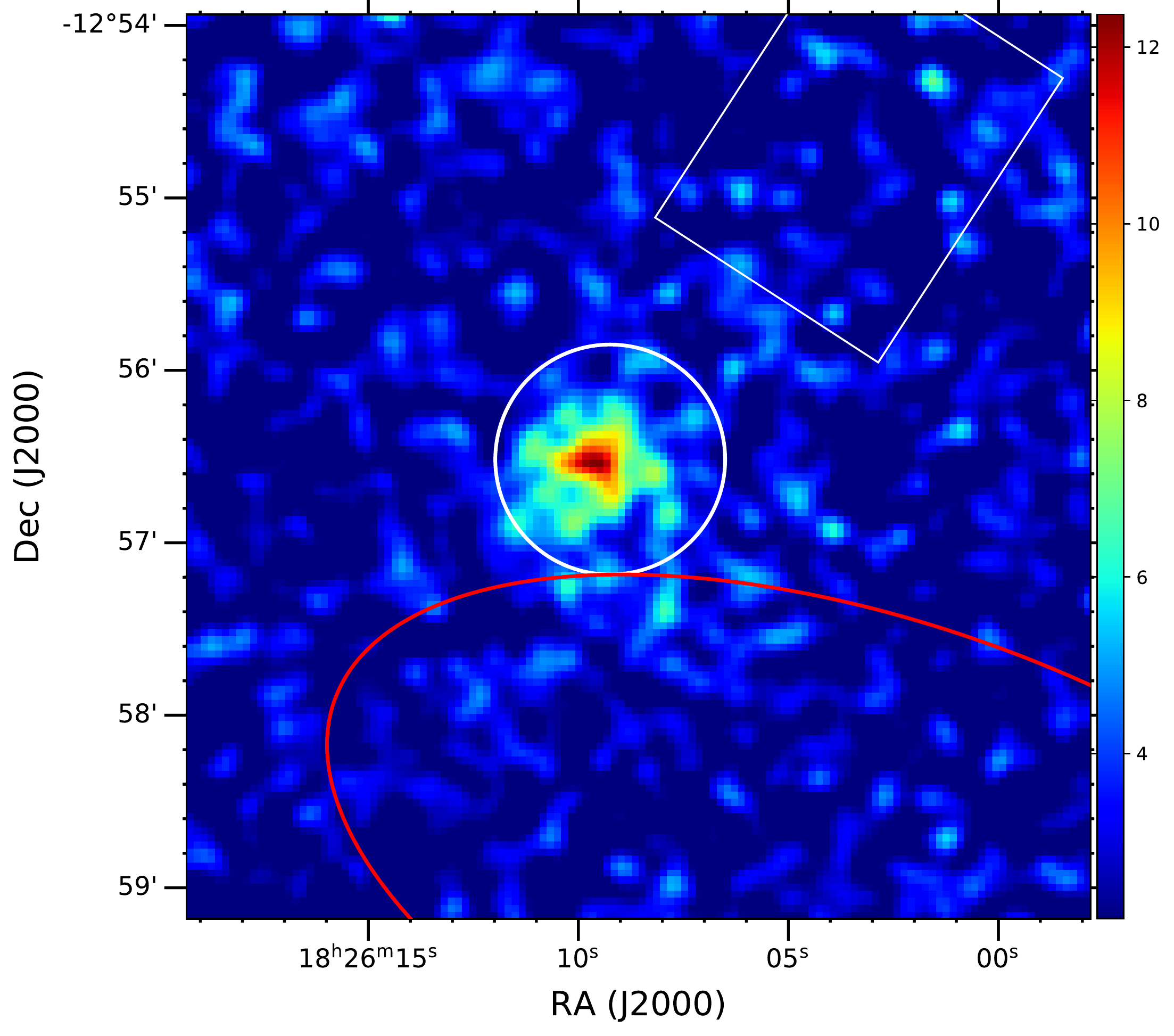} }}%
    \caption{\nustar\ images in the soft (3--10 keV; \textit{left panel}) and hard (10--20 keV; \textit{right panel}) energy bands. FPMA and FPMB images were merged after being smoothed with a Gaussian kernel ($\sigma = 3.69\asec$) and exposure corrected. Images are scaled to units of (exposure corrected) cts pixel$^{-1}$. 40\arcsec\ source extraction regions are shown in white and are adjacent to the rectangular background extraction regions. The red ellipse marks the diffuse PWN used in \xmm\ analysis and SED fitting.}%
    \label{fig:nustar}%
\end{figure*}

\subsection{NuSTAR}
\label{sec:nustar}

\nustar\ performed observations of \psr\ on October 25, 2019 (ObsID 40501006002). The total exposure time was 76.5 ksec. \nustar's field of view (FOV) of 13\amin$\times$13\amin\ frames both the pulsar and PWN on the sub-detector DET0 for both focal plane modules (FPMA and FPMB). The \nustar\ data were processed and analyzed with \texttt{NuSTARDAS Version 1.8.0} which is contained within \texttt{HEASOFT v6.27.2}, and we used the \nustar\  Calibration Database (CALDB) files \texttt{v20191219}.

\subsection{XMM-Newton}
\label{sec:xmm}

\xmmnewton\ performed a 140 ksec observation of \psr\ on October 11, 2014  (ObsID 0744420101). Data reduction and spectral extraction were performed in accordance with \texttt{XMM-SAS v18.0} procedures. This observation was performed with the EPIC-MOS (European Photon Imaging Camera Metal Oxide Semiconductor) CCD detectors MOS1 and MOS2 in full-frame mode with the medium optical filter, which frames the entire diffuse PWN. The pn CCD detector was operated in the Small Window mode (FOV of 4'$\times$4'), which is sufficient for pulsar and compact PWN analysis. Using calibrated event files, the data were reduced with the \texttt{emchain} and \texttt{epchain} tools. Using \texttt{espfilt}, good time intervals (GTIs) were found that excluded soft proton flares, which resulted in reduced exposure times of 77.4 ksec for MOS1, 80.8 ksec for MOS2, and 49.8 ksec for the pn. Canned FLAG values were used for MOS reduction, while FLAG was set to 0 for the pn. The corrected 0.5--10 keV images shown in Figure \ref{fig:xmm} were produced using the tasks \texttt{eimageget} and \texttt{eimagecombine}, with adaptive smoothing set to a radius between 1 and 4 pixels (1.1\arcsec\ per pixel).

\section{Radio observations}
\label{sec:radio}

Figure \ref{fig:radio} shows a 90 cm VLA field of the Eel PWN, containing numerous unidentified radio sources and large-scale emission across the region. The flux appears slightly increased near \psr\ and the diffuse Eel PWN, however due to the large-scale diffuse emission we were unable to obtain an accurate flux for the PWN alone. We derived an upper limit for the 90 cm PWN emission from the measured flux density in the diffuse PWN region (marked as a red ellipse in Figure \ref{fig:radio}). We use this upper limit of 1.89 Jy in our SED fits in \S\ref{sec:sed}-\S\ref{sec:evolutionary}. We estimate (using our evolutionary PWN modeling, see \S\ref{sec:evolutionary} with input parameters in Table \ref{tab:evolutionary_table}) that the diffuse Eel PWN's contribution to the 330 MHz flux density is 15.9 mJy, and at 1.4 GHz is 9.4 mJy. Whether the diffuse Eel PWN emission can be directly detected or isolated at 330 MHz or 1.4 GHz remains a question we will investigate in future work.
    
\begin{figure}[h!]
\begin{center} 
  \includegraphics[width=0.45\textwidth]{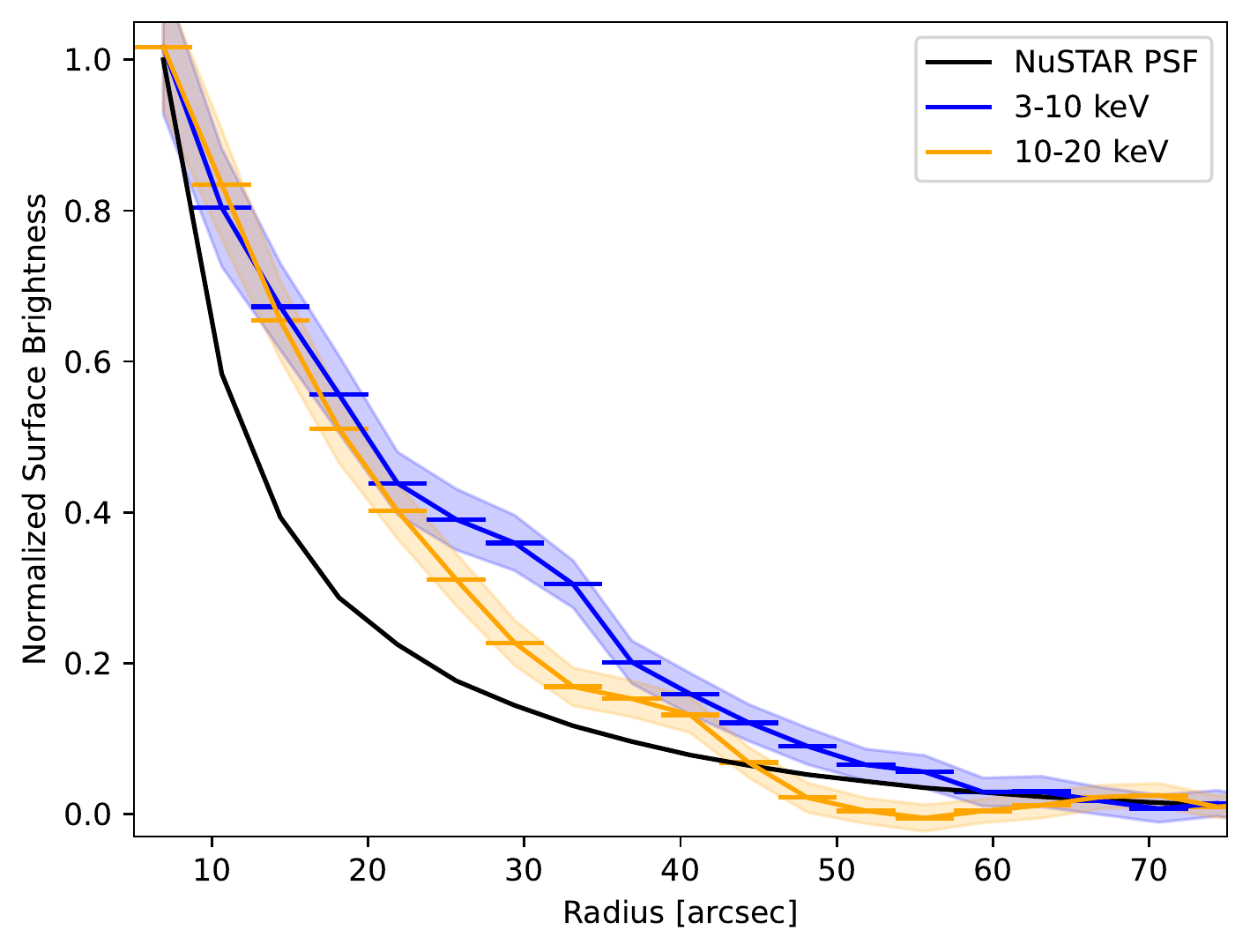}
  \caption{Radial profiles of the compact PWN and pulsar in the 3-10 keV (blue) and 10--20 keV (orange) bands, compared with the \nustar\ PSF (black). For a fair comparison, the radial profiles are normalized to 1 at the innermost radial bin (i.e. the pulsar position). The colored bands indicate 1$\sigma$ vertical errors, while the horizontal bars indicate radial bin sizes of 3.75\asec. The compact PWN is more extended in the soft band than in the hard band.}
  \label{fig:radprof}
\end{center}
\end{figure}

\section{\nustar\ Image Analysis}
\label{sec:imaging}

Previous X-ray studies of \psr\ \citep[e.g.;][]{Duvidovich2019, Karpova2019} with \xmmnewton\ have revealed spectral softening with distance from the pulsar, indicating evidence of the synchrotron burn-off effect.
In order to further study the energy dependence of the source extension, we took advantage of \nustar's broadband capabilities and analyzed images in the 3.0--10.0 (soft) keV  and 10.0--20.0 keV (hard) bands. The background was found to dominate above 20.0 keV.

We used the HEASoft \texttt{extractor} (v5.36) tool to filter the cleaned event files into the soft and hard bands.
Exposure maps were generated using the \texttt{nuexpomap} tool from \texttt{nupipeline} (v0.4.7), accounting for vignetting at 6.5 keV (soft images) and 15 keV (hard images).
The four event files (FPMA and FPMB in each energy band) and their exposure maps were imported into \texttt{XIMAGE} (v4.5.1).
Both the images and the associated exposure maps were smoothed using a $\sigma=3.69\asec$ (= 1.5 pixels) Gaussian kernel.
We then applied exposure correction by dividing each  source image by its exposure map.
Exposure corrected images from the FPMA and FPMB in each energy band were summed to generate mosaic images, after correcting for the relative offset between the FPMA and FPMB images by using the brightest pixel as a reference point. The processed images are shown in Figure~\ref{fig:nustar}.

To study source extension at different energies, radial profiles ($0\asec \leq r \leq 80\asec$) of the soft and hard images were generated around the brightest X-ray spot corresponding to the pulsar location.
Since the background was found to be spatially uniform near the source, the difference between the average flux and the average value of the \nustar\ point spread function (PSF) at $60\asec \leq r \leq 80\asec$ was subtracted from the radial profile to measure source extension.
A comparison with the on-axis \nustar\ PSF reveals extended emission from the compact PWN in both energy bands (Figure~\ref{fig:radprof}), with a significance $>5\sigma$ from $18\arcsec$--$35\arcsec$ in the soft band and a significance $>3\sigma$ from $10\arcsec$--$28\arcsec$ in the hard band. We also found that the soft image is more extended by $>3\sigma$ significance relative to the hard image from $25\arcsec$ to $35\arcsec$ (reaching a maximum difference of $5.4\sigma$ at $33\arcsec$), consistent with the synchrotron burn-off effect suggested by \cite{Duvidovich2019}.
The soft and hard band profiles follow each other up to $r \sim 25\asec$ likely because the X-ray emission in this region is dominated by the pulsar component. The dip in the hard band below the \nustar\ PSF at $45\arcsec$--$60\arcsec$ has less than a $2\sigma$ significance and is likely due to random fluctuations.

We further investigated 1-D spatial profiles in the \nustar\ images along two $20\asec$ wide axes, corresponding to ``jet'' and ``torus'' features reported by \cite{Karpova2019} using \chandra\ data.  We found an excess at 25\asec\ NE of the pulsar position in the soft band image (perhaps corresponding to the ``jet'' feature). However, this feature was detected only with a significance of $2\sigma$ likely because \nustar's PSF is too broad to resolve the PWN substructures detected by \chandra.

\section{X-ray Spectral Analysis}
\label{sec:xrayspec}

Here we extract and analyze spectra from the \nustar\ data alongside archived \xmmnewton\ data. We fit the \nustar\ and \xmmnewton\ spectra  jointly in order to constrain the pulsar emission from \psr\ and the compact Eel PWN. We then investigate the faint, diffuse Eel PWN to obtain X-ray data for SED fitting (see \S\ref{sec:sed}).

\subsection{The Compact Eel PWN}
\label{sec:compact_spec}

To perform hard X-ray spectral analysis on the compact Eel PWN, source extraction regions were determined with physical justification. \nustar's angular resolution (58\arcsec\ HPD, 18\arcsec\ full width half maximum) is limited in comparison to \xmmnewton's, therefore a larger extraction radius is required to account for the wider PSF. The bright compact PWN radius is 15\arcsec\ as measured by \chandra\ \citep{Karpova2019}, therefore we determined that a \nustar\ extraction radius of 40\arcsec\ maximized PWN counts while minimizing counts from the diffuse X-ray nebula and background. We did not find a significant difference between using 30\arcsec\ and 40\arcsec\ \nustar\ extraction radii for our source region in the compact PWN analysis. 

Annuli were avoided for background regions due to the diffuse PWN flux and high background levels in the \nustar\ data. Instead, we used source-free background regions on the same detector chip, which can be seen in Figure \ref{fig:nustar}. \nustar\ spectral extraction was performed with \texttt{nuproducts} and response files were generated for an extended source. The spectra were binned to 20 counts per bin. To obtain \xmm\ spectra, we used a 15\arcsec\ radius source spectral extraction region (the size of the bright, compact PWN) centered on \psr's location for MOS1, MOS2, and pn. We followed the standard \texttt{XMM-SAS v18.0} procedures, and auxiliary response files (ARFs) were generated for an extended source.

\begin{figure*}%
    \centering
    \subfloat {\includegraphics[width=0.71\columnwidth, angle=-90]{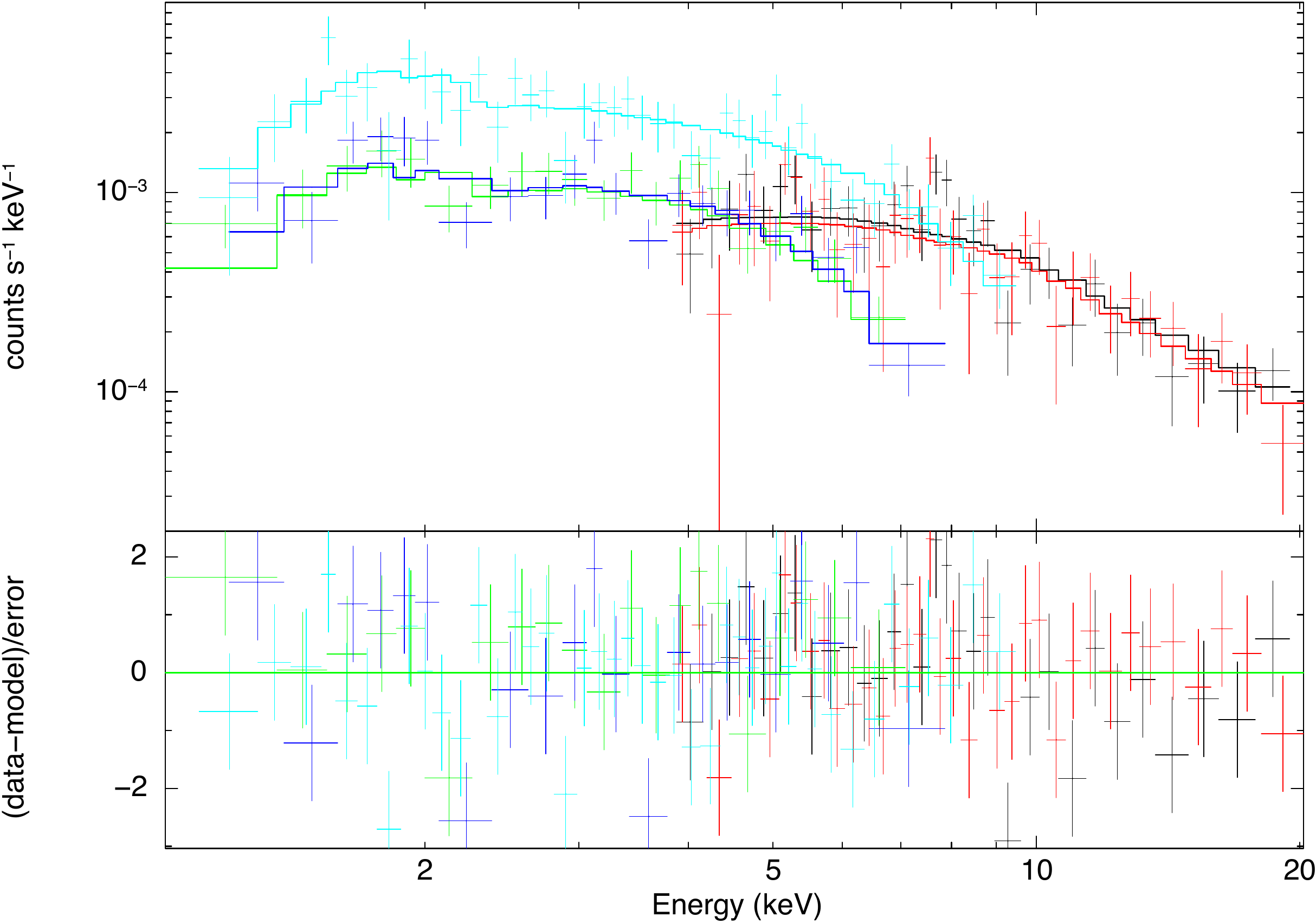}}%
    \qquad
    \subfloat {\includegraphics[width=0.71\columnwidth, angle=-90]{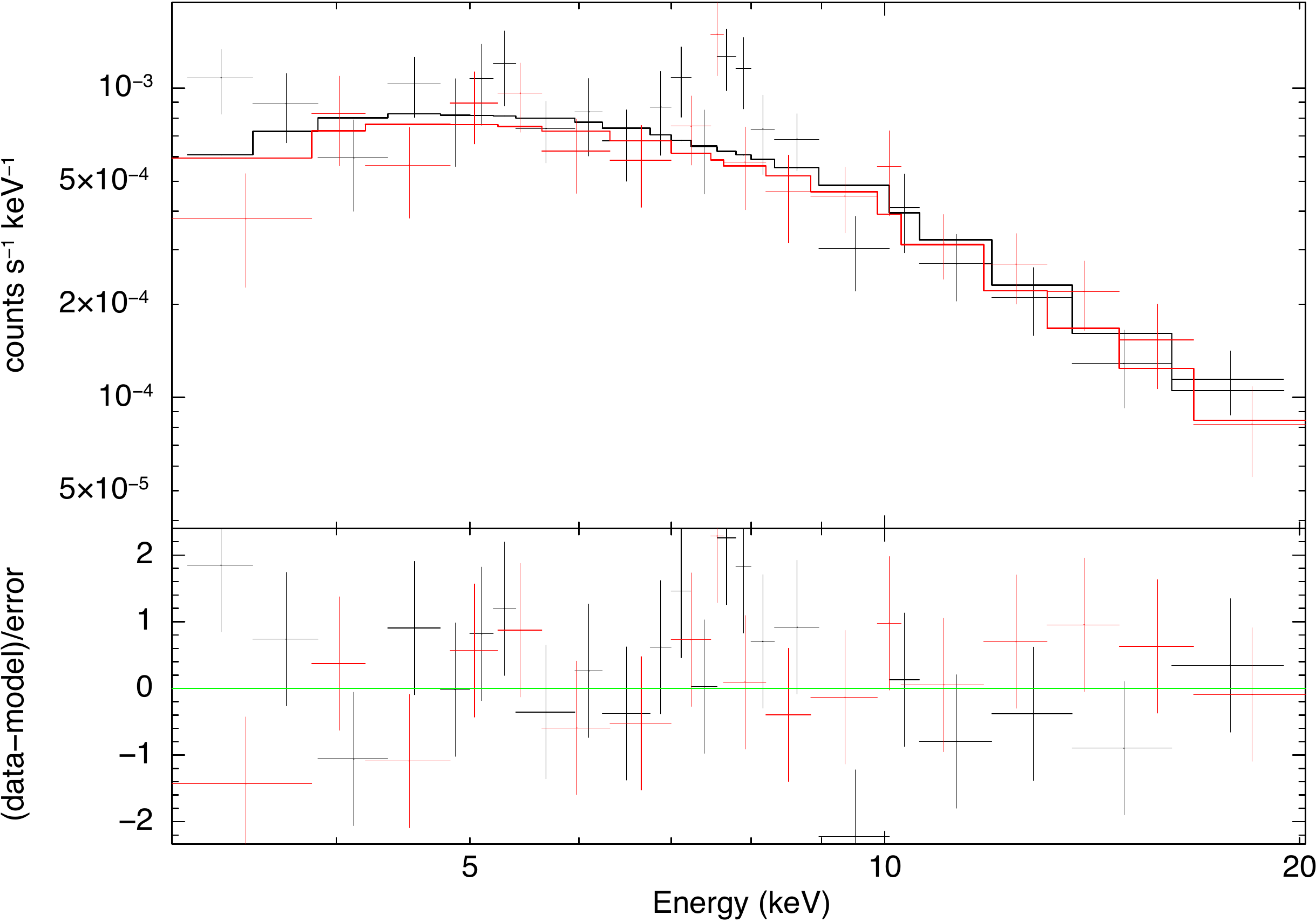}}%
    \caption{\textit{Left panel:} Compact PWN spectra including \nustar\ FPMA (black) and FPMB (red) from 4--20 keV, \xmmnewton\ MOS1 (dark blue) and MOS2 (green) from 1--8 keV, and \xmmnewton\ pn (light blue) from 1--10 keV. Each spectrum is grouped to minimum 20 counts per bin. The spectra are jointly fit by an absorbed power-law model. \textit{Right panel:} \nustar\ 3--20 keV spectra  (module A in black, module B in red), extracted from a $r=40$\arcsec\ circle around the pulsar and containing the compact PWN. Each spectral bin holds a minimum of 3$\sigma$ significance. The background was found to dominate in energies outside this range.}%
    \label{fig:xspec}%
\end{figure*}

Before fitting, the spectra were cut off above and below energies where the background began to dominate. For \nustar\ this gave a range from 4 to 20 keV, while the MOS1/2 spectra were fit between 1 to 8 keV and pn between 1 to 10 keV. For \nustar, this yielded 790 total and 562 net counts for FPMA, and 837 total and 475 net counts for FPMB. For \xmm, this yielded 403 total and 367 net counts for MOS1, 422 total and 380 net counts for MOS2, and 816 total and 655 net counts for pn. Using \texttt{XSPEC}, we initially fit the \nustar\ + \xmmnewton\ spectra to an absorbed  power-law model using a constant to normalize across the two telescopes: \texttt{const*tbabs*(powerlaw+powerlaw)}. Because neither \nustar\ nor \xmm\ can resolve the pulsar from the compact PWN, we added a second power-law to constrain the pulsar emission. Attempts to isolate the off-pulse emission in the \nustar\ observation using a timing solution were unsuccessful due to an unreliable $2\sigma$ pulsation detection. By freezing the pulsar's normalization and photon index, we were able to fit for the compact PWN's spectral parameters alone in both \nustar\ and \xmm. For the pulsar power-law component, we froze $\Gamma$ = 0.9 and a power-law normalization of $6.5 \times 10^{-6}$, both obtained from the spatially-resolved Chandra spectral analysis in \cite{Karpova2019}.
This fit (Figure \ref{fig:xspec}) yielded a photon index $\Gamma$ of 1.93 $\pm$ 0.19, a column density $N_H$ = (2.36 $\pm$ 0.34) $\times$ $10^{22}$ cm$^{-2}$, and a reduced $\chi^2$ = 1.07 for 151 degrees of freedom. The column density $N_H$ is consistent with \cite{Duvidovich2019} and \cite{Karpova2019} within error bars, while the photon index is slightly softer than the compact PWN photon index found by \cite{Karpova2019} using Chandra alone. All fitting parameters can be found in Table \ref{tab:xspecfit}. We did not find any obvious evidence of a spectral break in the \nustar\ data between 10--20 keV. The hard X-ray band spectra obtained by \nustar\ are shown in Figure \ref{fig:xspec}, with no significant residuals when fit to a power-law model (note the \nustar-only fit parameters can be found in Table \ref{tab:xspecfit}). 

\begin{figure}
\begin{center} 
  \includegraphics[width=0.85\columnwidth, angle=-90]{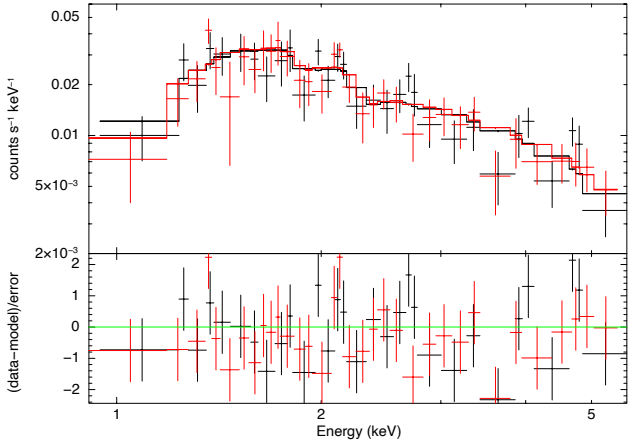}
  \caption{\xmmnewton\ MOS1 (black) and MOS2 (red) spectra extracted from  the diffuse PWN region (i.e. the red elliptical region in Figure \ref{fig:xmm}). Spectral fitting was performed in the  1--6 keV band, outside of which background dominates. Each bin in this plot holds a minimum of 3$\sigma$ significance.}
  \label{fig:ellipse_spectrum}
\end{center}
\end{figure}

\begin{table*}
\centering
\caption{\label{tab:xspecfit}
Pulsar-subtracted compact PWN and diffuse PWN X-ray spectral fitting parameters}
\begin{tabular}{ |c|c|c|c| }
    \hline
    Parameter & NuSTAR & NuSTAR + XMM & XMM-Newton \\
    \hline
    \hline
    Object & Compact PWN & Compact PWN & Diffuse PWN \\
    Region & 40\arcsec\ circle & 40\arcsec$_{NuSTAR}$ + 15\arcsec$_{XMM}$ & 3.5\arcmin\ $\times$ 1.6\arcmin\ ellipse \\
    Model & \texttt{tbabs*(po+po)} & \texttt{const*tbabs(po+po)} & \texttt{tbabs*powerlaw} \\ 
    $N_H$, $10^{22}$ cm$^{-2}$ & 2.36 (frozen) & 2.36 $\pm$  0.34 & 1.33 $\pm$  0.15 \\
    $\Gamma_X$ & 2.28 $\pm$ 0.26 & 1.93 $\pm$ 0.19 & 2.03 $\pm$ 0.15 \\
    PWN normalization$^a$ & 6.99 $\pm$ 3.35 & 3.70 $\pm$ 1.41 & 49.0 $\pm$ 9.3 \\
    cross-norm (MOS, pn)& --- & 0.69 $\pm$ 0.06, 0.65 $\pm$ 0.05 & --- \\
    $\chi^2_{\nu}$ (d.o.f) & 0.98 (73) & 1.07 (151) & 0.84 (155) \\
    flux, 0.5--10 keV$^b$ & 1.85$^{+0.06}_{-0.42}$ & 1.73$^{+0.04}_{-0.25}$ & 22.9$^{+1.9}_{-1.7}$ \\
    flux, 10--20 keV$^b$ & 1.65$^{+0.02}_{-0.13}$ & 1.77$^{+0.09}_{-0.13}$ & 1.33$^{+0.26}_{-0.22}$ \\
    \hline
    \multicolumn{4}{|c|}{$^a$norm in $10^{-5}$, $^b$unabsorbed flux in $10^{-13}$ erg s$^{-1}$ cm$^{-2}$} \\
    \hline
\end{tabular}
\end{table*}


\subsection{The Diffuse Eel PWN}
\label{sec:extended}

To investigate the properties of the diffuse, cometary Eel PWN in X-rays, we analyzed \xmmnewton\ and \nustar\ spectra of the diffuse PWN region (marked by a red ellipse in Figures \ref{fig:region}-\ref{fig:radio}). \xmmnewton\ analyses of this region were first reported by \cite{Duvidovich2019} and \cite{Karpova2019}. We followed the analysis procedures as used for the compact PWN in  \S\ref{sec:xmm} and \S\ref{sec:compact_spec}. The diffuse PWN region is defined as  an ellipse of radii 3.5\arcmin\ $\times$ 1.6\arcmin, angled to avoid pulsed emission from \psr. The \texttt{XMM-SAS v18.0} procedures were followed as before, and ARFs were generated for an extended source. The background was found to dominate outside of 1--6 keV, and fitting in this energy range yielded 8052 total and 4485 net counts for MOS1, and 8536 total and 4584 net counts for MOS2. The pn camera was operated in small-window mode for this observation and did not record counts from the entire diffuse PWN region. Fitting an absorbed power-law \texttt{tbabs*powerlaw} yielded a lower N$_H$ of 1.33 $\pm$ 0.15 $\times$ 10$^{22}$ cm$^{-2}$, a photon index $\Gamma$ = 2.03 $\pm$ 0.15, and a reduced $\chi^2$ = 0.84 for 155 degrees of freedom. The \xmmnewton\ MOS spectra are presented in Figure \ref{fig:ellipse_spectrum} after rebinning the data with a minimum $3\sigma$ significance in each bin. 

Since \nustar\ did not significantly detect the diffuse PWN, we determined a 68\% confidence interval upper limit for the diffuse PWN emission above 10 keV. We adopted only the FPMB data, since the FPMA data suffer from higher stray-light background contamination. We extracted \nustar\ FPMB spectra from the diffuse PWN region used for all \xmm\ analysis (the red ellipse in Figures \ref{fig:region}-\ref{fig:radio}) and fit an unabsorbed power-law model with $\Gamma =2.0$, the best-fit photon index of the diffuse PWN obtained by our fit to the \xmmnewton\ data (Table \ref{tab:xspecfit}). We raised the flux normalization to match the background flux levels measured by \nustar, yielding 10--20 keV and 20--30 keV flux upper limits of 1.0 $\times 10^{-12}$ erg cm$^{-2}$ s$^{-1}$ and 0.6 $\times 10^{-12}$ erg cm$^{-2}$ s$^{-1}$, respectively. These \nustar\ flux upper limits above 10 keV are presented in the subsequent SED plots along with \xmmnewton\ spectra of the diffuse PWN.

\section{Broadband PWN Spectral Modeling}
\label{sec:sed}

Here we present SED modeling of the TeV source data using three different models -- (1) a pure leptonic model, (2) a pure hadronic model, and (3) an evolutionary leptonic PWN model. In all sections, we adopted the \xmmnewton\ MOS1+2 spectra and \nustar\ hard X-ray flux upper limits of the diffuse PWN (the elliptical region shown in Figure \ref{fig:xmm}) as well as the \fermi\ upper limits at the TeV source position (\S\ref{sec:fermi}) and VLA 90 cm radio upper limit of the diffuse PWN region (\S\ref{sec:radio}). As shown in Table \ref{tab:xspecfit}, the diffuse PWN dominates in total X-ray flux, while the compact PWN's contribution is less than 10\% the diffuse PWN's flux \citep[or more precisely $\sim$4\% if the pulsar component is excluded;][]{Karpova2019}. In addition, the diffuse PWN is spatially coincident with the H.E.S.S. source, with the peak 5$\sigma$ flux of \hess\ aligned with a region of the diffuse PWN $\sim$6\amin\ offset from \psr\ (see H.E.S.S. contours in Figure \ref{fig:xmm}). For these reasons, we use diffuse PWN X-ray spectra in the following SED analysis. 

\hawc\ and \hess\ are offset in flux by a factor of $\sim$1.8--2, which could be explained by some combination of statistical and systematic errors, including both telescopes' modeling of the Galactic diffuse TeV emission \citep{Abdalla2021}. For this reason, and to avoid arbitrary scaling of one source's flux to the other, we fit our leptonic, hadronic, and evolutionary PWN models to the \hawc\ and \hess\ TeV datasets separately in order to explore and compare the resulting parameter spaces.

In all fits, we use \hawc\ and \hess\ spectral data from \citet{Albert2021} and \cite{Abdalla2020}, respectively. In \S\ref{sec:naima_leptonic}, we apply the \textsc{Naima} package for modeling SEDs \citep{Zabalza2015} to a pure leptonic scenario, considering the VLA 90 cm, \xmmnewton, \nustar, \fermi\ GeV, and HAWC/H.E.S.S TeV data. In \S\ref{sec:naima_hadronic}, we apply a pure hadronic model (also available in \textsc{Naima}) to the TeV data only, not using the X-ray data on account of their association with the (leptonic) diffuse PWN. In \S\ref{sec:evolutionary}, after ruling out the hadronic case and favoring a leptonic interpretation, we adopted a more sophisticated leptonic SED model that accounts for time evolution of the multi-wavelength PWN emission \citep{Gelfand2009}, jointly fitting VLA 90 cm, \xmm, \nustar, \fermi\ GeV, and HAWC/H.E.S.S TeV data. 

\subsection{Leptonic Model Fitting in \textsc{Naima}}
\label{sec:naima_leptonic}

A leptonic SED model generally consists of two primary components from radio to TeV energies. The lower-energy emission (up to $\sim$MeV) is created by synchrotron radiation of primary electrons accelerated in the PWN. The GeV and TeV $\gamma$-ray emission is produced by Inverse Compton scattering (ICS) of the CMB and ambient photons by the same electrons. In addition, the model contains a synchrotron self-Compton (SSC) component which takes into account  synchrotron photons up-scattered by energetic electrons inside the PWN. However, the SSC component is usually less significant than the ICS component in all but the most compact nebulae.

We constructed our leptonic SED model using the \texttt{InverseCompton} and \texttt{Synchrotron} components in \textsc{Naima}. We assumed an exponential cut-off power-law (ECPL) model for the electron energy distribution ($f(E) \propto AE^{-\alpha}e^{-E/E_{cut}}$ electrons cm$^{-3}$ erg$^{-1}$).
\textsc{Naima}'s SED data fitting is performed using a  Markov Chain Monte Carlo (MCMC) algorithm which samples posterior distributions of a set of model parameters, iteratively optimizing their respective $\chi^2$ values. The free parameters that we fit to VLA, \xmm, \nustar, \fermi, and either HAWC or H.E.S.S. data are a power-law index $\alpha$, cut-off energy, flux normalization of the PWN electron energy distribution, PWN B-field, and FIR photon energy density. \cite{Voisin2016} investigated the ISM in this region and analyzed CO lines and maser emission consistent with molecular clouds. Assuming the large optical depth limit for the $^{12}$CO line, we derived a low excitation temperature range of $T_{exc}$ $<$ 20 K from the antenna temperatures reported for the nearby CO emission region labeled as ``R1c" \cite{Voisin2016}. We assumed  $T = 15$ K as the temperature of the local FIR photon field and left its energy density as a free parameter. 

\begin{figure}
\begin{center} 
  \includegraphics[width=1.0\columnwidth]{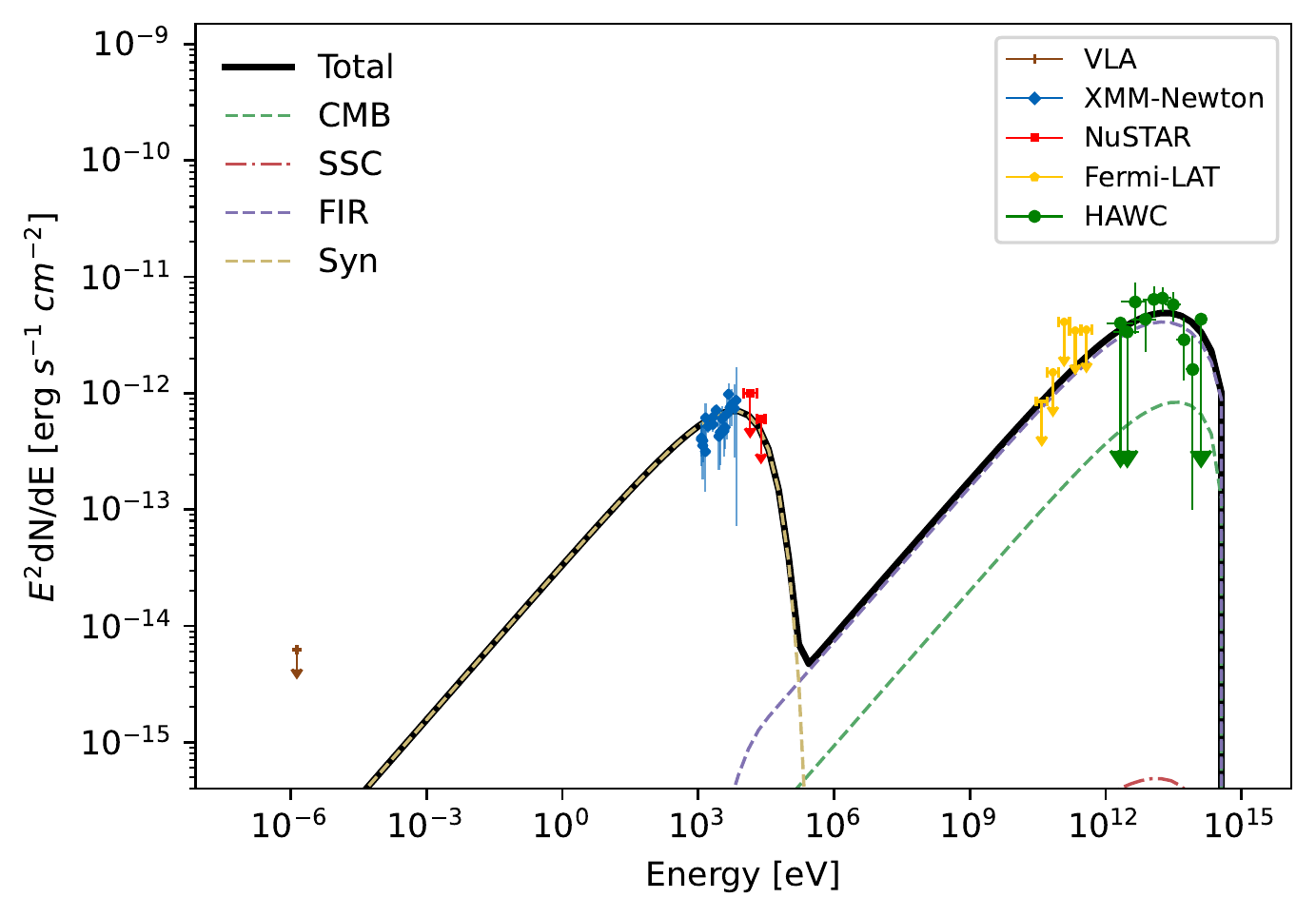}
   \caption{A \textsc{Naima} leptonic model fit to the broadband SED with radio, X-ray, GeV upper limits, and \hawc\ spectral data. The electron energy distribution is assumed to be an exponential-cutoff power-law model. The fit parameters can be found in Table \ref{tab:sedfit}.}
  \label{fig:hawcleptonic}
\end{center}
\end{figure}

\begin{figure}
\begin{center} 
  \includegraphics[width=1.0\columnwidth]{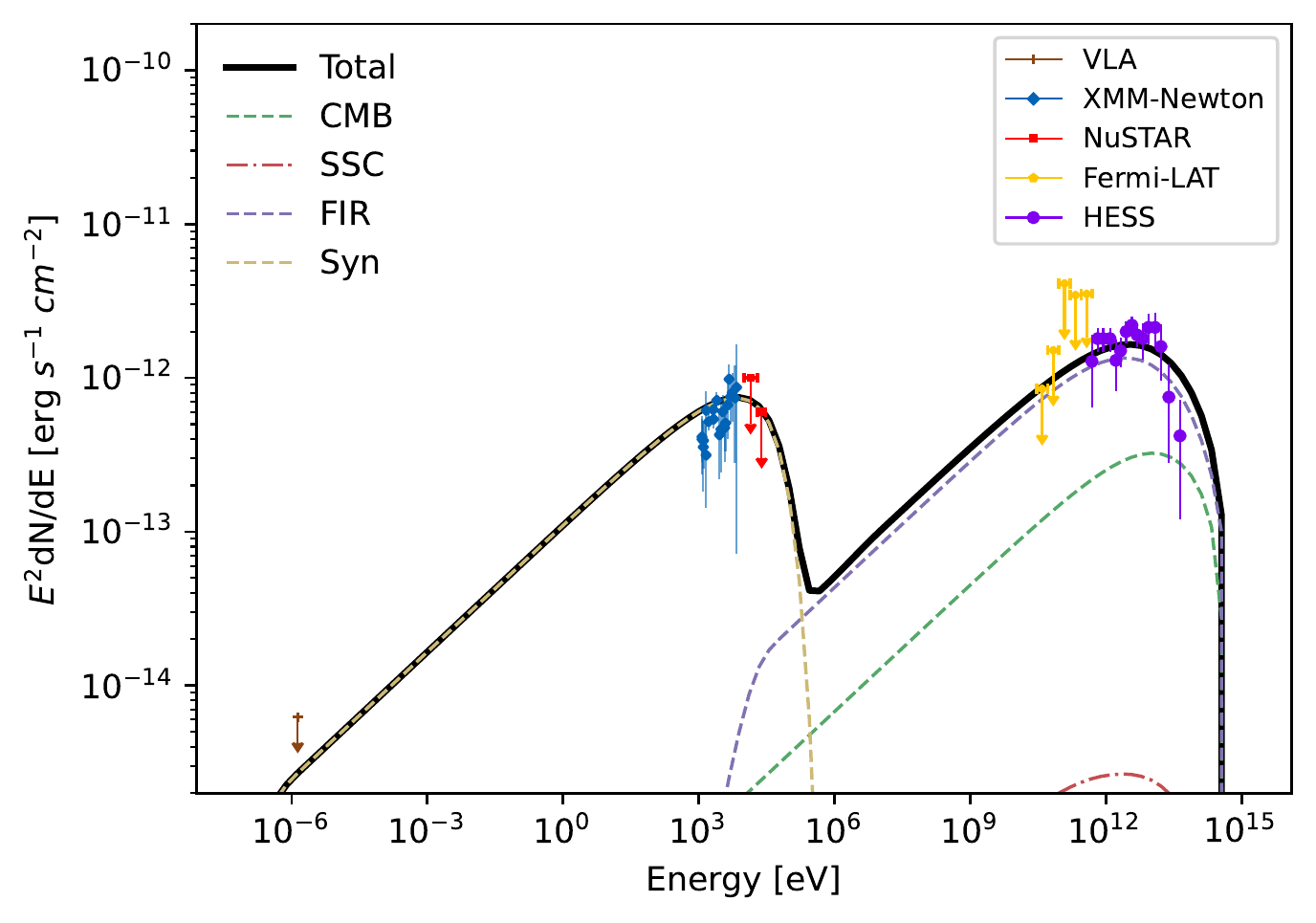}
  \caption{A \textsc{Naima} leptonic model fit to  the broadband SED with radio, X-ray, GeV upper limits, and \hess\ spectral data. Though the HAWC and H.E.S.S. observations likely detected the same source, their $\gamma$-ray fluxes are slightly offset from one another, potentially as a result of systematics or differing analyses between the two telescopes \citep{Abdalla2021}. The fit parameters can be found in Table \ref{tab:sedfit}.}.
  \label{fig:hessleptonic}
\end{center}
\end{figure}

\begin{figure}
\begin{center} 
  \includegraphics[width=1.0\columnwidth]{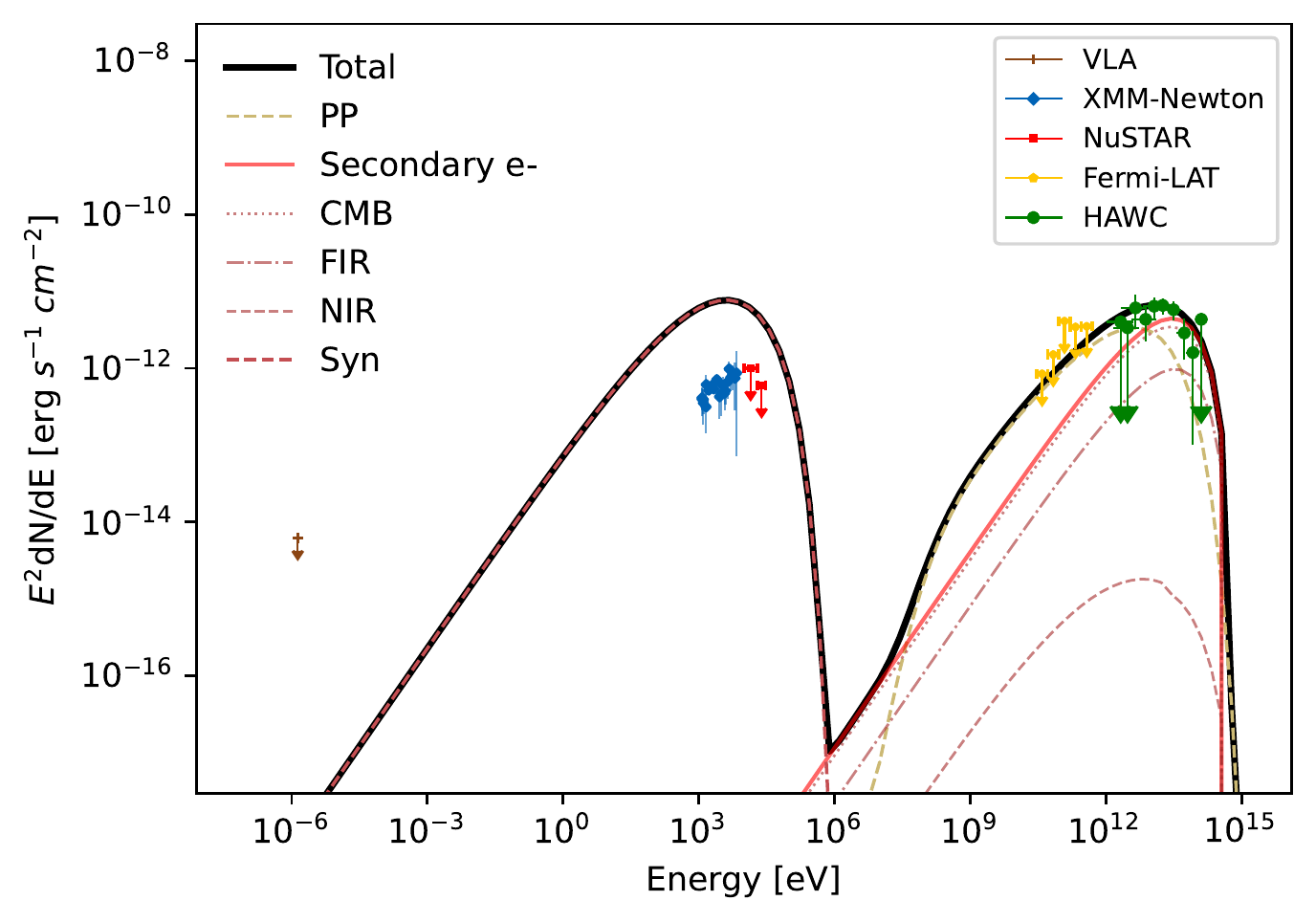}
  \caption{An example of the \textsc{Naima} hadronic model fit to the \fermi\ upper limits and \hawc\ spectral data. The X-ray spectral data demonstrate the overprediction of observed X-ray flux as a result of fixing the $W_e/W_p$ ratio to 1/3 (see \S\ref{sec:naima_hadronic}). In this fit, $W_e$ = 3.3 $\times 10^{46}$ erg, $W_p$ = 1 $\times 10^{47}$ erg, power-law index $\alpha$ = 1.3, cut-off energy = 100 TeV, $n_H = 600$ cm$^{-3}$ and B = 3 $\mu$G.}
  \label{fig:hawchadronic}
\end{center}
\end{figure}

When we fit the HAWC data alongside the X-ray data, we found that the best-fit model yields a power-law index of $\alpha = 2.2$ and a cut-off energy of $E_{cut} = 590$ TeV. The best-fit B-field is low at $1.9$ $\mu$G, and the regional FIR energy density is elevated at $4.4$ eV cm$^{-3}$ in order to match the TeV fluxes. The total electron energy injected into the PWN is $W_e = 1.8 \times 10^{46}$ erg. The best fit to the \hess\ TeV data yields a slightly softer power-law index $\alpha = 2.5$ as well as the same cut-off energy $E_{cut} = 590$ TeV. The B-field is slightly higher at $3.0$ $\mu$G, and the resulting FIR energy density is lower at $2.9$ eV cm$^{-3}$. The total electron energy required in this model is $W_e = 1.6 \times 10^{46}$ erg. Our SED data and best-fit models are shown in Figures \ref{fig:hawcleptonic}-\ref{fig:hessleptonic}, and all fit parameters are listed in Table \ref{tab:sedfit}. The similarity between SED fitting parameters combined with their spatial association strongly suggests that \hawc\ and \hess\ are two separate detections of the same TeV $\gamma$-ray accelerator, as previously noted by \cite{Albert2021}. Given the successful fitting with this generic, leptonic model, we performed a deeper investigation with a more complex, time-evolution leptonic PWN model in \S\ref{sec:evolutionary}. 

\begin{table}[h!]
\caption{Fit parameters obtained from the leptonic  \textsc{Naima} model}
\label{tab:sedfit}
\centering
    \begin{tabular}{ |c|c|c| }
    \hline
    Parameter & \hawc\ & \hess \\
    \hline
    B--field & 1.9 $\mu$G & $3.0$ $\mu$G \\
    power-law index $\alpha$ & $2.2$ & $2.5$ \\
    Cut-off energy & $590$ TeV & $590$ TeV \\
    FIR field & $4.4$ eV cm$^{-3}$ & $2.9$ eV cm$^{-3}$\\
    FIR temperature & 15 K & 15 K \\
    Electron energy & $1.8 \times 10^{46}$ erg & $1.6 \times 10^{46}$ erg \\
    \hline
    \multicolumn{3}{|c|}{E$_{min}$= 0.5 GeV} \\
    \hline
    \end{tabular}
\centering
\end{table}

\subsection{Hadronic Model Fitting in \textsc{Naima}}
\label{sec:naima_hadronic}

In the hadronic case, relativistic proton collisions with the ISM produce both neutral and charged pions. Energetic neutral $\pi^0$ mesons can decay into photons above $\sim$100 TeV, making hadronic processes a plausible mechanism for some PeVatron candidates \citep[such as SNR G106.3+2.7;][]{Amenomori2021}. Additionally, the charged pions decay into secondary electrons and positrons, which contribute to ICS $\gamma$-ray emission and synchrotron radiation at X-ray energies and below. Since the X-ray emission around the pulsar is most likely associated with leptonic processes in both the compact and diffuse PWN regions, we do not include the X-ray data obtained by \xmm\ and \nustar\ in our hadronic model fit. Instead, the observed X-ray emission can serve as an upper limit for the diffuse synchrotron X-ray emission expected from hadronic secondary electrons or hadronic thermal X-ray emission.


To investigate the hadronic scenario, we fit the hadronic \texttt{PionDecay} model, available in the  \textsc{Naima} package, to both the HAWC and H.E.S.S. TeV data. We assumed an ECPL model for the proton energy distribution  ($f(E) \propto AE^{-\alpha}e^{-E/E_{cut}}$ protons cm$^{-3}$ erg$^{-1}$). As a target for the hadronic  collisions, we considered the CO emission region at 3.5 kpc distance \cite{Voisin2016} labeled as ``R1c", a kinematic velocity component of region ``R1" which is consistent with the distance to \psr\ found by \cite{Karpova2019}. We derived a mean hydrogen density $n_H$ = 600 cm$^{-3}$ from the $n_{H_2}$ = 220 cm$^{-3}$ virial mass of the CO line emission component and the relation $n_H$ = 2.8 $n_{H_2}$ \citep{Voisin2016}. 
The best-fit model using the \hawc\ data yields a power-law index of $\alpha = 1.5$ and a cut-off energy of $E_{cut}$ = 100 TeV. The total proton energy required in this model is $W_P = 1.0 \times 10^{47}$ erg. The \hess\ TeV data can be fit with the same power-law index of $\alpha = 1.5$ and a lower cut-off energy $E_{cut}$ = 40 TeV due to the lack of higher energy bins. The total proton energy required in this model is $W_P = 3.0 \times 10^{46}$ erg. Note that \textsc{Naima} does not model the secondary electron energy distribution self-consistently from   the proton-proton collisions in the \texttt{PionDecay} model component. Hence, following the method of \cite{Mori2020}, we added separate \texttt{Synchrotron} and \texttt{InverseCompton} components to account for the contribution of secondary electrons in the X-ray and TeV bands. We assumed the same energy spectrum for protons and secondary electrons (electron $\alpha = 1.5$ and cut-off energy $E_{cut}$ = 100 TeV for \hawc). We adjusted the flux normalizations of the synchrotron and ICS components by imposing that the total energy of secondary electrons ($W_e$) is 1/3 of the total energy of injected protons, which accounts for the branching ratios of charged pions produced from p-p collisions \citep{Coerver2019, Mori2020}.

As can be seen in Figure \ref{fig:hawchadronic}, we found that the predicted synchrotron emission component from the secondary electron population, obtained by fixing the $W_e/W_p$ ratio to 1/3, vastly overpredicts the PWN X-ray flux even if a low ISM B-field of 3 $\mu$G \citep{Crutcher2012} is assumed. In this model, any B-field $>$ 1 $\mu$G overpredicts the X-ray flux, while B $<$ 1 $\mu$G is unable to match the observed X-ray spectral shape. The lack of this predicted diffuse X-ray emission in the TeV source region (which would appear substantially higher in flux than the compact or diffuse PWN at typical ISM B-fields), combined with the presence of an energetic pulsar and diffuse PWN coincident with \hawc\ and \hess, rules out the pure hadronic case as the sole origin of the TeV emission.

\begin{figure}
\begin{center} 
  \includegraphics[width=1.0\columnwidth]{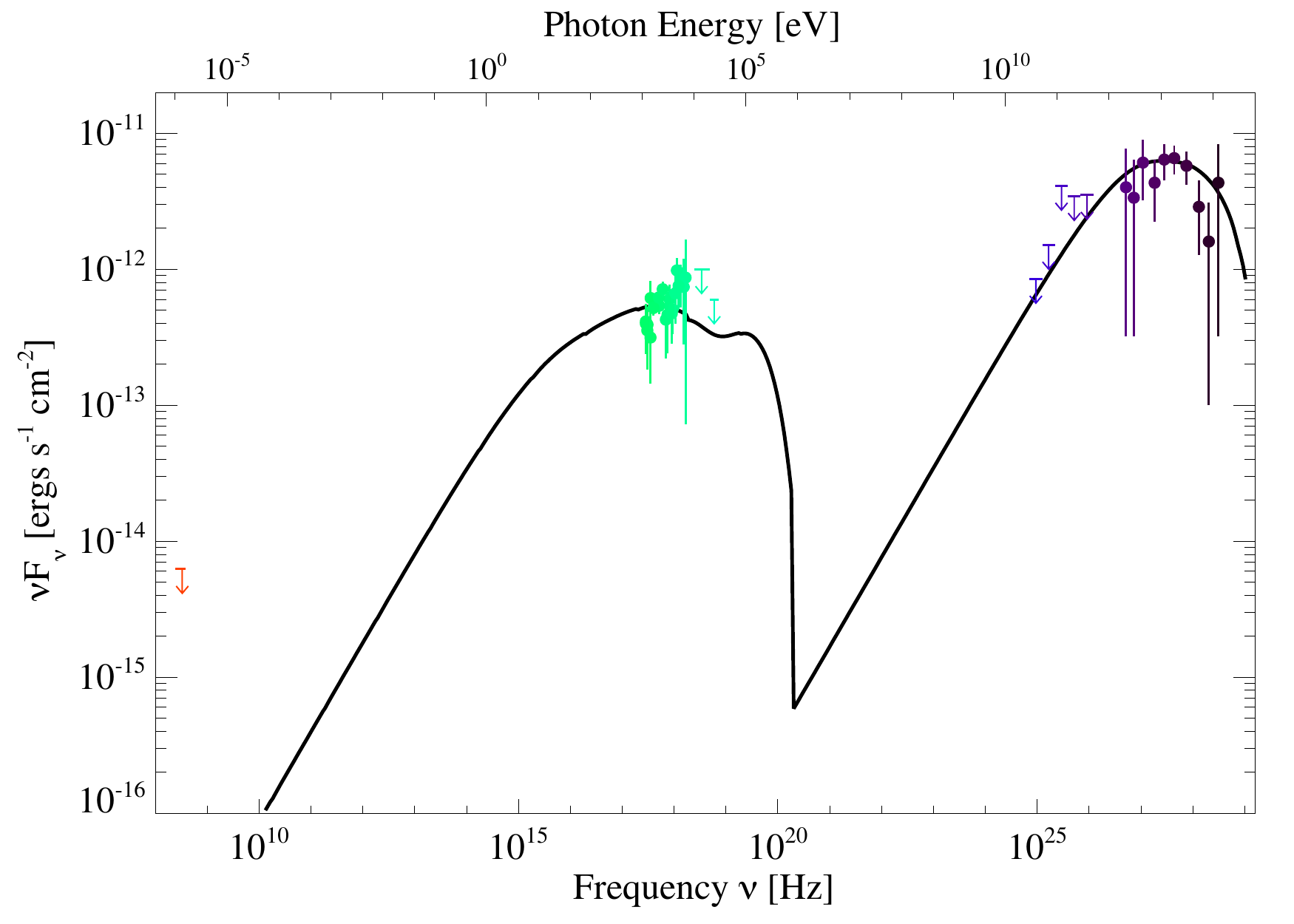}
  \caption{An evolutionary PWN model fit of the broadband SED with \hawc\ TeV data (dark red) and \fermi\ upper limits (purple), along with \xmmnewton\ X-ray data, \nustar\ upper limits (green), and a VLA 90cm upper limit (red). The fit parameters can be found in Table \ref{tab:evolutionary_table}.}
  \label{fig:hawcevolutionary}
\end{center}
\end{figure}

\begin{figure}
\begin{center} 
  \includegraphics[width=1.0\columnwidth]{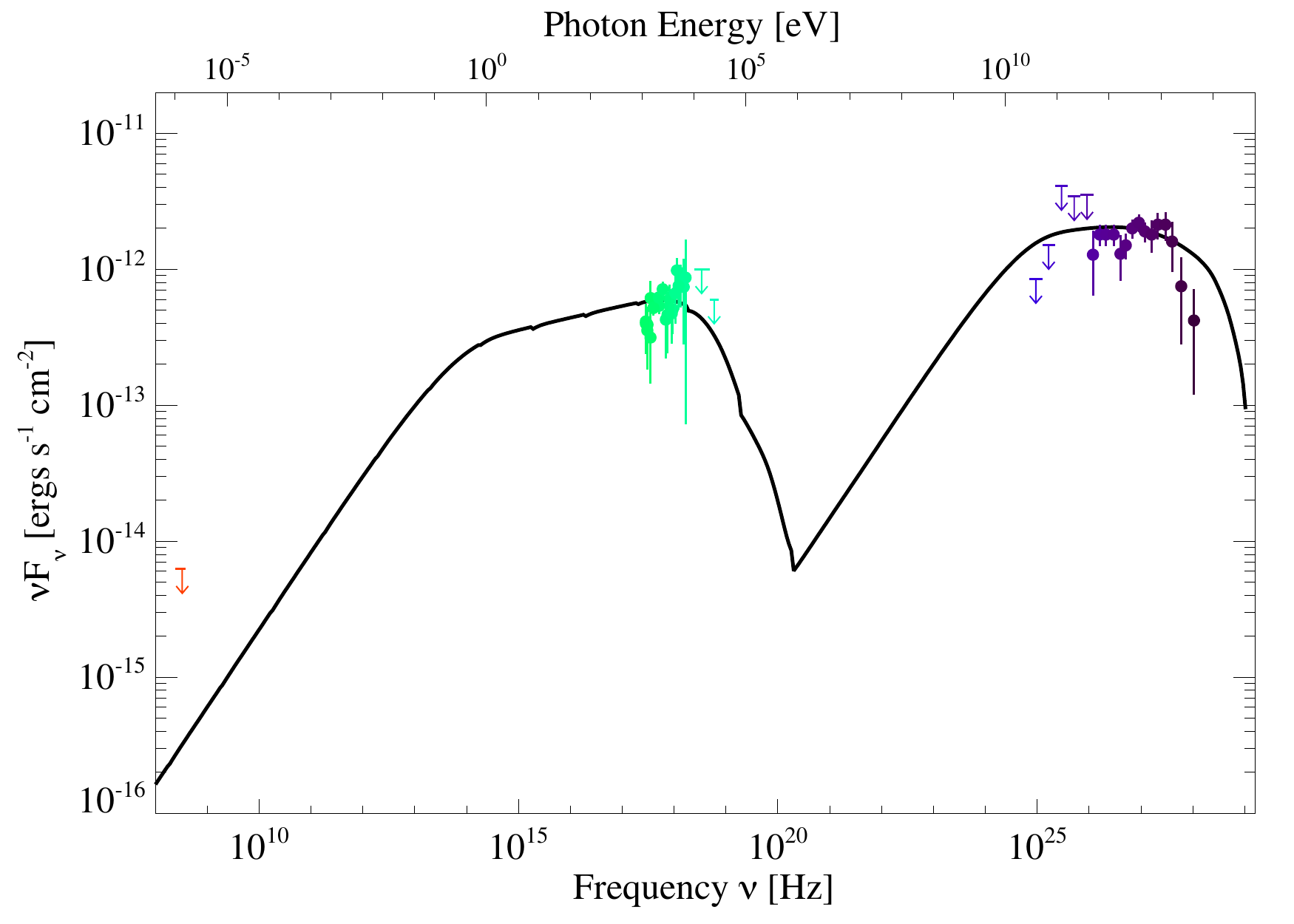}
  \caption{An evolutionary PWN model fit to the broadband SED with \hess\ data (dark red), \fermi\ upper limits (purple), \xmmnewton\ X-ray data, \nustar\ upper limits (green), and a VLA 90cm upper limit (red). The fit parameters can be found in Table \ref{tab:evolutionary_table}.}
  \label{fig:hessevolutionary}
\end{center}
\end{figure}

\subsection{Evolutionary PWN Model}
\label{sec:evolutionary}

In order to estimate the physical parameters of the progenitor supernova, associated neutron star, and its pulsar wind in the system, we fit the observed properties of the diffuse Eel PWN with the predictions of a physical model for the dynamical and radiative evolution of a spherical PWN inside a SNR (see \cite{Gelfand2009} for a detailed description of this particular model, and \cite{Gelfand2017} for a review of similar dynamical, radiative PWN models).

In order to obtain posterior minimum $\chi^2$ parameters for the Eel PWN, we made certain assumptions within this model to account for uncertain or unconstrained features of the pulsar, SNR and ISM. Firstly, we assumed that the unshocked SN ejecta density profile features a uniform density core which is surrounded by an envelope with $\rho \propto r^{-9}$. This is a common assumption, but may not be fully representative of the varied ejecta density profiles of various progenitor SNRs \citep{Gelfand2017}. Next, we assumed that the supernova ejecta with initial kinetic energy $E_{sn}$ and mass $M_{ej}$ is expanding into a medium with uniform density $n_{ism}$. Using the observed current spin-down luminosity $\dot{E}$ and characteristic age $t_{ch}$ of \psr\ as inputs for the model (and ensuring that each trial combination of input parameters reproduced these observables), for a given braking index $p$ and spin-down timescale $\tau_{sd}$ we determined the system's true age $t_{age}$ and initial spin-down luminosity $\dot{E}_0$ as follows:
\begin{eqnarray}
    t_{age} & = & \frac{2t_{ch}}{p-1} - \tau_{sd} \\
    \dot{E}_0 & = & \dot{E} \left(1 + \frac{t_{age}}{\tau_{sd}} \right)^{\frac{p+1}{p-1}}.
\end{eqnarray}

Next, we assumed that the entire spin-down luminosity $\dot{E}$ of \psr\ is injected into the PWN as either magnetic fields $\dot{E}_B$ or the kinetic energy $\dot{E}_p$ of relativistic leptons:

\begin{eqnarray}
    \dot{E}_B(t) & = & \eta_B\dot{E}(t) = \eta_B \dot{E}_0 \left(1 + \frac{t}{\tau_{sd}} \right)^{-\frac{p+1}{p-1}} \\
    \dot{E}_p(t) & = & (1-\eta_B)\dot{E}(t) = (1-\eta_B)\dot{E}_0\left(1 + \frac{t}{\tau_{sd}} \right)^{-\frac{p+1}{p-1}}
\end{eqnarray}

holding $\eta_B$ constant with time. Next, we assumed that the spectrum of particles injected into the PWN can be described by a broken power-law:

\begin{equation}
\frac{d\dot{N}}{dE} = \begin{cases} \dot{N}_{break}\left(\frac{E}{E_{break}}\right)^{-p_1} E_{min} < E < E_{break} \\ \dot{N}_{break}\left(\frac{E}{E_{break}}\right)^{-p_2} E_{break} < E < E_{max} \end{cases}
\end{equation}

where we further assumed that the free parameters $E_{min}$, $E_{break}$, $E_{max}$, $p_1$ and $p_2$ are constant with time. The power-law normalization $\dot{N}_{break}$ across all times $t$ is calculated by the relation:

\begin{equation}
\dot{E}_p = \int_{E_{min}}^{E_{max}}E\frac{d\dot{N}}{dE}dE.
\end{equation}

Lastly, we assumed that synchrotron and
inverse Compton scattering are the only two mechanisms for electron radiative losses within the PWN, and that the B-field across the diffuse Eel PWN has a uniform strength \citep[see][]{Gelfand2009}.

The total electron energy, magnetic field strength, and radius of the SNR reverse shock $R_{RS}$ are output by the model. All input and output parameters for this model are listed in Table \ref{tab:evolutionary_table}. No priors were used in our model fitting -- the possible value distribution for all parameters is assumed to be unbounded and completely flat (with the exception of the wind magnetization, which as a fraction necessarily must be $<$ 1). A Metropolis Markov Chain Monte Carlo (MCMC) algorithm (see \cite{Gelman2013} and Section 3.2 of \cite{Gelfand2015} for more details) was used to determine the combination of parameters which best reproduced the observed X-ray and $\gamma$-ray fluxes of the diffuse Eel PWN, the observed radius of the coincident radio SNR candidate G18.45$-$0.05 \citep[7.0\amin\ $\pm$ 0.6\amin;][]{Anderson2017}, and a circularly-approximated PWN radius. As a best estimate for the full size of the $\gamma$-ray emitting PWN, we used our measured 3$\sigma$ extension of the H.E.S.S source at E $>$ 10 TeV (Figures \ref{fig:region} and \ref{fig:xmm}) of 6\amin\ or $\sim$6 pc at $D = 3.5$ kpc in all model fits, with $\pm 1$ pc uncertainty.

The combinations of input parameters which resulted in the lowest $\chi^2$ (\hawc: $\chi^2$ = 22.85 for 22 degrees of freedom and \hess: $\chi^2$ = 33.6 for 27 degrees of freedom) are given in Table \ref{tab:evolutionary_table}. As shown in Figures \ref{fig:hawcevolutionary} and \ref{fig:hessevolutionary}, for this set of model parameters, inverse Compton scattering CMB photons alone is sufficient to explain the observed $\gamma$-ray emission. The addition of a low energy density photon field is compatible with this set of model parameters but did not improve the quality of the fit. 

Several SNR/PWN systems have been found with comparable non-canonical parameters (e.g. our predicted progenitor SN explosion energy and ejecta mass, see Table \ref{tab:evolutionary_table}). For example, \cite{Guest2019} derived a SN explosion energy of 3 $\times$ 10$^{49}$ erg in their analysis of the G21.5$-$0.9 progenitor SNR, and \cite{Gotthelf2021} found a low ejecta mass of 0.51 $M_\odot$ and a SN explosion energy of 1.26 $\times$ $10^{50}$ erg in their application of \cite{Gelfand2009}'s evolutionary model to the SNR Kes 75. Additionally, it has been suggested that the Crab progenitor SN was of an unusually low energy \citep[$\sim$10$^{50}$ erg;][]{Yang2015}. A pulsar braking index $>$ 3 as suggested by our fit to \hawc\ is unusual, but such braking indices have been measured for anomalous pulsars such as PSR J1640$-$4631 and could be explained by invoking higher order multipoles \citep{Archibald2016}. The \hess\ fit braking index is of a typical magnitude, $\leq$ 3. 

As detailed by \cite{Gelfand2015} in their analysis of G54.1+0.3, there can be significant degeneracies between the different parameters of this model, for example between the mass and initial kinetic energy of the supernova ejecta with the density of the surrounding ISM, and between the braking index and spin-down timescale of the pulsar. Further parameter exploration is needed to derive proper statistical bounds on the values of these quantities. Further investigation of the degeneracy between braking index and spin-down timescale may allow for lower braking indices when fitting the HAWC spectrum. The pulsar spin-down timescale in our modeling is higher than most other systems, with best-fit values of 8.4 and 9.8 kyr for the HAWC and H.E.S.S. sources, respectively. However, other time-evolution fits to PWNe have yielded high spin-down timescales -- for example, \cite{Torres2014} modeled a spin-down age of 6.6 kyr in a fit to HESS J1356–645, an offset relic TeV PWN, with a pulsar of similar \edot\ and comparable PWN morphology to the Eel. 

The ISM density, \hess\ fit pulsar braking index, and wind magnetization parameters all agree with previous applications of time-evolution PWNe models \citep{Torres2014, Gelfand2015}. The low and maximum energy particle indices are also in agreement with the single power-law index found by \textsc{Naima}. Furthermore, the spherically-approximated PWN radius ($6$ pc) and SNR ($\sim$8--10 pc) predicted by this set of model parameters agree well with the observed PWN volume and SNR candidate radius.

\begin{table*}
\caption{Minimum $\chi^2$ PWN and SNR model parameters for the Eel PWN, using \nustar, \fermi, and \textit{VLA} upper limits} fit to \xmmnewton, \hawc, and \hess\ data
\centering
\label{tab:evolutionary_table}
    \begin{tabular}{ |c|c|c| }
    \hline
    Model Parameter & \hawc\ & \hess\ \\
    \hline
    SN explosion energy & $1.4 \times 10^{50}$ ergs & $3.5 \times 10^{49}$ ergs \\
    SN ejecta mass & 1.3 $M_\odot$ & 1.4 $M_\odot$ \\
    ISM density & 0.2 cm$^{-3}$ & 0.2 cm$^{-3}$ \\
    Total electron energy & $3.5\times10^{47}$~ergs & $5\times10^{47}$~ergs \\
    Magnetic field strength & 0.6 $\mu$G & 1.5 $\mu$G \\
    PWN radius $R_{PWN}$ & 5.95 pc & 6.08 pc \\
    SNR forward shock radius $R_{FS}$ & 10.1 pc & 8.34 pc \\
    SNR reverse shock radius $R_{RS}$ & 5.97 pc & 6.00 pc \\
    Pulsar braking index & 3.2 & 2.8 \\
    Pulsar spin-down timescale & 8.4 kyr & 9.8 kyr \\
    True age & 4600 yr & 6700 yr \\
    Initial spin-down luminosity & $8 \times 10^{36}$ ergs/s & $1.1 \times 10^{37}$ ergs/s \\
    Wind magnetization & 0.001 & 0.005 \\
    Min. injected particle energy & 0.3 GeV & 0.5 GeV \\
    Max. injected particle energy & 2.3 PeV & 1.6 PeV \\
    Break injected particle energy & 28 TeV & 3.4 TeV \\
    Low energy particle index & 1.7 & 1.9 \\
    High energy particle index & 2.6 & 2.8 \\
    \hline
    \multicolumn{3}{|c|}{For $\dot{E} = 3.6 \times 10^{36}$ ergs/s and CMB-only photon field} \\
    \hline
    \end{tabular}
\end{table*}

\section{Discussion}
\label{sec:disc}

Our multi-wavelength SED and morphology study establishes that \hawc\ and \hess\ are associated with the GeV pulsar PSR~J1826$-$1256 and its Eel pulsar wind nebula, confirming similar suggestions by \cite{Abdalla2020}, \cite{Duvidovich2019}, \cite{Karpova2019}, \cite{Anguner2017} and \cite{Roberts2007}. \hawc\ and \hess\ likely represent independent observations of the same TeV source \citep[this paper;][]{Albert2021} and its TeV emission (as resolved by H.E.S.S. above 10 TeV) spatially coincides with the diffuse X-ray PWN. 
Both the \textsc{Naima} leptonic SED model (assuming a homogeneous population of relativistic electrons; \S\ref{sec:naima_leptonic}) and the evolutionary PWN model (\S\ref{sec:evolutionary}) yield fit parameters that are consistent with the PWN hypothesis. However, in comparison with other middle-aged PWNe, there are several distinct features of the Eel PWN which may account for its energetic TeV emission as a possible PeVatron candidate. Based on the pulsar and PWN parameters determined by our SED analysis, we investigate the properties of the Eel PWN in this section. 

\subsection{PWN Energetics}

While our independent leptonic SED models \citep[\textsc{Naima} and][]{Gelfand2009} yielded similar particle indices and low B-fields, they differ in the best-fit total electron energy and photon field energy density values. This is primarily because our evolutionary PWN model takes into account electron injection and cooling throughout the modeled pulsar true age, while the generic \textsc{Naima} model assumes no time evolution of the electron energy distribution. 
The total electron energy ($W_e = $ (1.6--1.8) $\times$ $10^{46}$ ergs) obtained by the \textsc{Naima} model is an order of magnitude lower than that of the evolutionary PWN model ($W_e =$ (3.5--5.0) $\times$ $10^{47}$ ergs). As a result, the ICS component in the \textsc{Naima} model requires an elevated FIR photon field energy density of 2.9--4.4 eV cm$^{-3}$ (in addition to the CMB component) to fit the observed TeV flux.  

However, the \textsc{Naima} results are likely incompatible with the observed TeV $\gamma$-ray extension of \hess. Note that the electron cooling time due to pitch angle averaged synchrotron and inverse Compton energy losses in the Thomson regime can be expressed by: \citep{Rybicki1979}
\begin{equation}
    \tcE \approx\ 24.5\left(1 + \frac{U_{ph}}{U_B}\right)^{-1} \gamma_7^{-1} \hspace{0.1cm} B_5^{-2} \hspace{0.1cm} {\rm  kyr},
\end{equation} 

where $U_{ph}$ and $U_B$ are the photon field and magnetic field energy densities, respectively, $\gamma_7 = E/(10^7m_e c^2)$ is the particle Lorentz factor, and $B_5 = B \times 10^5$ G is the magnetic field. Therefore, given the FIR energy density and PWN B-field ($B= 1.9$--3.0 $\mu$G) required by the \textsc{Naima} model, $\simgt$40 TeV electrons (which would emit 10 TeV ICS $\gamma$-rays in these conditions) will cool in $<$ 2 kyr, preventing them from diffusing across the entire observed extension of \hess\ (see \S\ref{sec:bfield}, (8) and (9)). 
On the other hand, in the case of the  evolutionary PWN model, the cooling time of $\sim$60 TeV electrons (which would emit $\gamma$-rays above 10 TeV in a CMB-only photon field) is far greater than our modeled pulsar true age (4.6--6.7 kyr). This suggests that the injected electrons in our evolutionary PWN model should be able to emit TeV $\gamma$-rays throughout the true pulsar age. Dividing the total electron energy of (3.5--5.0) $\times$ $10^{47}$ ergs by the spin-down power ($\dot{E} = 3.6 \times 10^{36}$ erg s$^{-1}$) yields an injection timescale of 3.1--4.4 kyr which is comparable to our modeled true age. In the case of a cooling timescale $>$ true age and injection timescale $\sim$ true age as suggested by the evolutionary PWN model parameters, 4.6--6.7 kyr also provides sufficient time for these electrons to diffuse across the observed TeV $\gamma$-ray extension of \hess\ (\S\ref{sec:bfield}).    
Therefore, in addition to the fact that the evolutionary PWN model is more physically motivated, we consider it plausible that the FIR photon field is not significant in the Eel PWN region and that the electron energy is on the order $\sim$10$^{47}$ ergs -- these assumptions are adopted in the subsequent B-field analysis (\S\ref{sec:bfield}).

To find the total rotational energy released from the pulsar ($E_{\rm tot}$) we integrated the spin-down power (\edot) over the true age suggested by our evolutionary model ($t$ = 4.6--6.7 kyr). Next, dividing the total electron energy $W_e$ by $E_{\rm tot}$ indicated the fraction of rotational energy that was spent injecting electrons during the PWN evolution. The evolutionary model fits to the \hawc\ and \hess\ data predicted an electron energy of (3.5--5.0) $\times$ $10^{47}$ ergs, and \edot\ integrated over true age is (5.2--7.6) $\times$ $10^{47}$ ergs, implying that approximately two-thirds of the total rotational energy has been injected into the PWN.

Given the pulsar's spin-down power ($\dot{E} = 3.6 \times 10^{36}$ erg s$^{-1}$), we found that the 0.5--8 keV X-ray and 1--10 TeV TeV efficiencies of the Eel PWN are $\eta_{keV}$ = 5.5 $\times$ 10$^{-4}$ and $\eta_{TeV}$ = (2--4) $\times$ 10$^{-3}$ respectively, assuming a pulsar distance of 3.5 kpc. The TeV efficiency range reflects the different values for TeV luminosity obtained by the HAWC and H.E.S.S source data. 
These X-ray and TeV efficiencies are consistent with those of middle-aged PWNe  \citep{Kargaltsev2015}. Additionally, \cite{Karpova2019} found that the TeV and X-ray luminosity and $L_{TeV} / L_{keV}$ ratio versus system age were consistent with the luminosity and ratio-to-age distributions of other PWNe detected in the TeV band \citep{Abdalla2018, Kargaltsev2013}.

\subsection{Comparison with other middle-aged PWNe } 

While the PWN hypothesis is well-supported, the exact properties of the diffuse Eel PWN remain uncertain. These uncertainties include the mechanism responsible for the highly asymmetrical X-ray morphology and the magnitude of the ambient B-field inside this extended X-ray and $\gamma$-ray tail. One possible explanation could be that the Eel PWN is produced by a supersonic pulsar, which has been suggested due to the PWN's diffuse X-ray morphology and its similarity in appearance to other supersonic PWNe \citep{Roberts2007, Kargaltsev2017}. Another possible explanation could be an ongoing collision with the SNR reverse shock, which would prevent the outflow of radiating particles near the shock front and allow all remaining particles to diffuse away freely in the opposite direction, creating an asymmetric PWN which is often found to be offset from the pulsar position \citep{Blondin2001, Karpova2019}. 

Table \ref{tab:eelcomparison} shows a comparison of the Eel PWN to three other evolved PWNe which are also characterised by a compact X-ray nebula around the pulsar and diffuse X-ray emission. MSH 15$-$56 and G327.1$-$1.1 are both thought to have interacted with the SNR reverse shock, resulting in soft X-ray thermal line emission and radio brightness, while the origin of CTB 87's asymmetric  morphology remains uncertain \citep{Guest2020}. Unlike the other three evolved PWNe, the Eel PWN lacks any significant radio emission \citep[a signature often seen in PWNe interacting with SNR reverse shocks;][]{Blondin2001}, with our estimated 1.4 GHz flux density just 0.04$\%$ of CTB 87 and even less for MSH 15$-$56 and G327.1$-$1.1. Considering our modeled true age of 4.6--6.7 kyr (Table \ref{tab:evolutionary_table}) compared to 17--20 kyr in the other PWNe, the Eel may not have had sufficient time to undergo substantial reverse shock interaction. 

In addition to the lack of radio emission around the Eel PWN, our one-zone evolutionary model in its spherical approximation of the PWN and SNR also suggests that the PWN-reverse shock interaction has either not yet begun, or is just starting (see Table \ref{tab:evolutionary_table}). While the pulsar's proper motion may cause one side of the PWN to interact with the reverse shock first and eventually result in an asymmetrical morphology, we did not find any significant evidence of thermal emission in our compact PWN spectra (see Figure \ref{fig:xspec}). Instead, the compact X-ray nebula is characterised by non-thermal X-ray emission with evidence of the synchrotron burn-off effect. The lack of features such as thermal emission lines in the compact PWN, an X-ray bright SNR, and a bright relic radio PWN coincident with the diffuse Eel PWN \citep[all features seen in the reverse shock-interacting PWNe MSH 15$-$56 and G327.1$-$1.1;][]{Temim2013, Temim2015} alongside the uncertain results of our modeled reverse shock prediction suggest that the Eel PWN may not have yet commenced a substantial interaction with the reverse shock of its host SNR. Therefore, the supersonic PWN hypothesis remains a plausible explanation for the asymmetric, diffuse X-ray emission seen extending from \psr.

\begin{table*}
\centering
\caption{\label{tab:eelcomparison}
Comparison of the Eel PWN to other offset, evolved PWNe with similar X-ray morphologies}
\begin{tabular}{ |c|c|c|c|c| }
    \hline
    Parameter & Eel & MSH 15--56$^a$ & G327.1--1.1$^b$ & CTB 87$^c$ \\
    \hline
    \hline
    $\Gamma_X$, compact PWN & 1.2$^{+0.24}_{-0.23}$ $^d$ & 1.41$^{+0.24}_{-0.24}$ & 1.61$^{+0.08}_{-0.07}$ & 1.76$^{+0.04}_{-0.03}$  \\
    $\Gamma_X$, diffuse PWN & 2.03$^{+0.15}_{-0.15}$ $^e$ & 1.84$^{+0.13}_{-0.11}$ & 2.11$^{+0.04}_{-0.05}$ & 1.91$^{+0.08}_{-0.07}$ \\
    Flux, compact PWN & 8.5$^{+1.0}_{-0.9} \times 10^{-14}$ $^d$ & 0.52 $\times  10^{-13}$ & 0.45 $\times 10^{-12}$ & 0.59 $\times 10^{-12}$ \\
    Flux, diffuse PWN & $2.3^{+0.2}_{-0.2} \times 10^{-12}$ $^e$ & 3.2 $\times 10^{-13}$ & 3.68 $\times 10^{-12}$ & 1.60 $\times 10^{-12}$ \\
    F$_{PWN,compact}$ / F$_{PWN,diffuse}$& 3--4.48\% & 16.3\% & 12.2\% & 36.5\% \\
    PWN radio flux density & 0.009 Jy (1.4 GHz)$^f$ & 26 Jy (1.0 GHz) & 2.1 $\pm$ 0.4 Jy (1.5 GHz)$^g$ & 9 Jy (1 GHz) \\
    System age & $\sim$6 / 14.4 kyr$^h$  & 16.5 kyr (SNR) & 17.4 kyr (SNR) & 20 kyr$^i$ \\
    \hline

    \multicolumn{5}{c}{\parbox{12cm}{\textbf{Notes:} $^a$\cite{Temim2013}, 0.3--10 keV flux in erg cm$^{-2}$ s$^{-1}$; $^b$\cite{Temim2015}, 0.3--10 keV flux in erg cm$^{-2}$ s$^{-1}$; $^c$\cite{Guest2020}, 0.3--10 keV flux in erg cm$^{-2}$ s$^{-1}$; $^d$\cite{Karpova2019}, 0.5--10 keV flux in erg cm$^{-2}$ s$^{-1}$; $^e$this paper, see \S\ref{sec:extended}; $^f$this paper, estimate from evolutionary PWN model; $^g$\cite{Ma2016}; $^h$modeled true age vs pulsar characteristic age; $^i$relic PWN age}} \\
\end{tabular}
\end{table*}


\subsection{Magnetic Field of the Diffuse PWN}
\label{sec:bfield}

The magnetic field strength B inside the diffuse Eel PWN (the location of a majority of the TeV-emitting electrons) has not been previously well constrained. \cite{Duvidovich2019} suggested that B $> 2$ $\mu$G based on the peak energy of the diffuse X-ray spectrum and the assumption that the synchrotron cooling time must be less than the pulsar's characteristic age. \cite{Karpova2019} suggested B $\sim5$ $\mu$G on the basis of particle diffusion lengths across the angular size of \hess. Similarly, we find that low  B-fields ($\sim$0.7--3 $\mu$G) are required to jointly fit the X-ray and TeV data in both the \textsc{Naima} leptonic SED and evolutionary PWN models (see Tables \ref{tab:sedfit} and \ref{tab:evolutionary_table}). Low B-fields have previously been found in other TeV-detected PWNe: fitting a different evolutionary PWN model to multi-wavelength SED data of G292.2$–$0.5 and HESS J1356$–$645 yielded B = 4 $\mu$G and 3.1 $\mu$G, respectively \citep{Torres2014}. Low wind magnetization fractions (we obtained typical values in our evolutionary model fits of 0.1-0.5\%, see Table \ref{tab:evolutionary_table}) indicate that only a small fraction of the pulsar's spin-down luminosity is injected into the PWN in the form of magnetic fields. Additionally, adiabatic expansion of the PWN decreases its magnetic field strength, as shown by \cite{Gelfand2009}, \cite{Tanaka2010}, and other similar works. A lower PWN B-field increases the synchrotron lifetime of the diffusing particles, which is consistent with the large angular extensions of the Eel PWN in the X-ray and TeV bands.  Our \textsc{Naima} parameters are derived from generic nonthermal particle modeling that does not account for time-evolution or cooling. For this reason, below we investigate the cooling and diffusion of electrons using the PWN parameters obtained by the evolutionary PWN model. For all transverse distances and diffusion lengths, we assumed that the pulsar distance is 3.5 kpc. 

While synchrotron X-ray radiation of leptons in the diffuse PWN falls below \xmmnewton\ sensitivities at distances greater than 4--6\amin\ from the pulsar, this X-ray faint region is spatially associated with the brightest $\gamma$-rays from \hess\ (see the offset location of the \hess\ $5\sigma$ peak flux, Figure \ref{fig:xmm}). This offset between the diffuse PWN detected by \xmmnewton\ and $\gamma$-ray centroid indicates that the true extension of the diffuse PWN is likely larger than 4--6\amin, with some PWN particles reaching the \hess\ peak 6--8\amin\ from the pulsar. This suggests long diffusion lengths for ICS leptons in the PWN. Additionally, the \textit{ASCA} source catalog in 2001 reported a diffuse X-ray source extended over 8\amin\ from \psr\ \citep{Roberts2001}.
As a best estimate for the diffusion lengths required for particles to fill the observed TeV source at a given energy, we used our measured 3$\sigma$ extension of the H.E.S.S source at E $>$ 10 TeV (Figures \ref{fig:region} and \ref{fig:xmm}) of 6\amin\ or $\sim$6 pc at $D = 3.5$ kpc. Given that the H.E.S.S source spectrum peaks at $E\sim10$ TeV, we assumed that the ICS $\gamma$-ray energy at this radius is typically $E_{\rm ICS} \sim 10$ TeV. 

CMB photons alone are sufficient to account for the TeV emission in our evolutionary PWN model. In a CMB-only photon field, 63 TeV electrons would emit 10 TeV $\gamma$-rays via ICS. These $\sim$60 TeV electrons have a cooling time (7) of 15 kyr for the derived B-field of $\sim$1 $\mu$G. 
Next, we calculated how far these energetic electrons can diffuse within their cooling time with this formula: 
\begin{equation}
    R_{\rm d} = 2\sqrt{D_B(t_{\rm d})},
    \label{eq:rdif}
\end{equation}
where $t_{\rm d}$ is the diffusion timescale after escape and $R_{\rm d}$ is a lower limit for diffusion length. Assuming Bohm-type diffusion in the region surrounding \psr, the diffusion coefficient \db\ can be written as 
\begin{equation}
    D_B = \frac{Ec}{3qB} \sim 3.33 \times 10^{26}\: \text{cm}^2 \text{s}^{-1}
    \left(\frac{E}{100 \hspace{0.1cm} \text{TeV}} \right)
    \left(\frac{B}{10 \mu \text{G}} \right)^{-1}.
    \label{eq:bohm}
\end{equation} 

Using our evolutionary PWN model parameters (Table \ref{tab:evolutionary_table}), 63 TeV electrons within B = 0.6--1.5 $\mu$G have a cooling time ($\sim$15 kyr) greater than our modeled true pulsar age of 4.6--6.7 kyr. Using this true age to calculate a diffusion radius instead of the cooling time, the 63 TeV electrons are able to diffuse to 6 pc, which is consistent with the \hess\ extension above 10 TeV. 

If the pulsar is fast moving, its proper motion could modify the size of the X-ray and TeV emission region. However, we find that the pulsar will travel between 2--7 pc within our estimated true pulsar age, assuming a transverse velocity between 500--1,000 km\,s$^{-1}$. In either case, the estimated diffusion length (plus an additional distance due to the pulsar's proper motion) is comparable to the measured 3$\sigma$ extension of \hess\ above 10 TeV. Therefore, our estimated PWN B-field of $\sim$1 $\mu$G is well justified due to its consistency with both the broadband SED data and the TeV source size.

Our estimated diffuse PWN B-field is also compatible with estimations of electron diffusion in the X-ray band. The peak of the diffuse PWN's synchrotron X-ray spectrum (1.8 keV, see Figure \ref{fig:ellipse_spectrum}) indicates that 210 TeV electrons will radiate 1.8 keV photons in a $\sim$1 $\mu$G B-field, with electron energy inversely proportional to B-field. These electrons will cool over a timescale longer than our estimated true pulsar age in a CMB-only environment, therefore using the true age (4.6--6.7 kyr) as a diffusion timescale suggests that electrons radiating at $\sim$2 keV will be able to diffuse $>$ 6 pc in a B-field of $\sim$1 $\mu$G, which is consistent with both the broadband SED modeling and the observed diffuse PWN size in X-rays. 

\subsection{The Eel as a PeVatron PWN?}

In fitting our evolutionary PWN model to the \hawc\ and \hess\ data over a large parameter space, we found that a maximum electron energy $E_{\rm max} \simgt 1$ PeV is strongly preferred with the observed \psr\ spin-down power of $\dot{E}  =  3.6 \times 10^{36}$ erg\,s$^{-1}$ as a fixed input. This is because a lower $E_{\rm max}$ requires a large fraction of the spin-down power to be injected into lower energy electrons, thus forcing our time-evolution SED model to over-predict the observed $\gamma$-ray fluxes at $E < 10$ TeV and the GeV upper limits. 
The observed TeV $\gamma$-ray fluxes and GeV upper limits combined with the pulsar's high spin-down power make a robust case for $E_{\rm max} \simgt 1$ PeV in our multi-wavelength evolutionary SED modeling of the Eel PWN, and parameter exploration with $E_{\rm max} < 1$ PeV did not yield a successful fit. Even without time dependence, the generic \textsc{Naima} leptonic model also suggests high electron energies, requiring a cut-off electron energy of $E_{\rm max} \sim 600$~TeV. 
While there are some  uncertainties associated with our model assumptions (e.g. representing an asymmetrical PWN as spherical), both the leptonic \textsc{Naima} model and our evolutionary PWN model suggest the acceleration of leptons to energies on the order of hundreds of TeV up to a maximum of $\sim$2 PeV.

\section{Conclusion}
\label{sec:conclusion}

Recent $\gamma$-ray and X-ray observations have suggested that the Eel PWN is a counterpart to the bright HAWC source \hawc, also detected by H.E.S.S primarily above $E\sim1$ TeV. The HAWC detection up to nearly $E\sim100$ TeV suggests that the TeV source may be a PeVatron candidate, i.e., a long-sought, extreme particle accelerator in our galaxy \citep{Albert2021}. \nustar\ has revealed evidence of strong synchrotron burn-off in the compact X-ray nebula up to $\sim$20 keV, while a previous \xmm\ observation detected tail-like diffuse X-ray emission extending $\sim$6$\amin$ from the pulsar. The diffuse X-ray PWN coincides well with the H.E.S.S TeV source detected above 10 TeV, while the contribution of the compact PWN and the pulsar is less than 10\% in the X-ray band. 

Based on our multi-wavelength SED and morphology study, we ruled out a pure hadronic origin of the TeV emission due to the inability of a hadronic SED model to fit both the X-ray and TeV data (along with the spatial coincidence between the diffuse Eel PWN and \hess\ above 10 TeV). Through our time-evolution leptonic SED modeling as well as theoretical estimates of the electron cooling and diffusion timescales, we found that (1) the radiative efficiencies in the X-ray and TeV bands are comparable to those of other middle-aged PWNe; (2) the true pulsar age is $\sim$5.7 kyr (4.6--6.7 kyr) which is significantly smaller than the spin-down age of 14.4 kyr; (3) the PWN B-field is low at $B\sim$ 1 $\mu$G; (4) the asymmetric X-ray nebula morphology is likely due to supersonic motion of the pulsar; (5) the lack of radio emission is notable compared to other evolved PWNe with similar X-ray and TeV properties; (6) the maximum electron energy may extend above $\sim$1 PeV. Our modeled maximum electron energy reaching $\sim$1 PeV and beyond is particularly exciting, as it further suggests the Eel PWN may be a leptonic PeVatron. 

Why the Eel PWN stands out as an extreme leptonic particle accelerator will remain an elusive question until other TeV-detected PWNe are further investigated in similar multi-wavelength approaches. These investigations are ongoing with seven other middle-aged PWNe (including several PeVatron candidates) detected by HAWC and LHAASO \citep{Mori2021}. In a larger context, it is still unclear whether evolved PWNe such as the Eel are dominant contributors to the Galactic PeVatron population as well as the positron excess \citep{Malyshev2009}. As demonstrated in this paper, an extensive X-ray survey of middle-aged PWNe with  \nustar, \xmm\ and other X-ray telescopes, along with use of next-generation TeV telescopes such as the Cherenkov Telescope Array (CTA) and the Southern Wide-field Gamma-ray Observatory (SWGO),  will be essential to answering these fundamental questions in high-energy astrophysics. 

\facilities{NuSTAR}

\software{HEAsoft (v6.22), NuSTARDAS (v1.8.1), XMM-SAS (v18.0), XSPEC (v12.9.1; \cite{Arnaud1996}), Naima \citep{Zabalza2015}, Fermipy \citep{Wood2017}. This research made use of APLpy, an open-source plotting package for Python \citep{Robitaille2012}.}

\begin{acknowledgments}
The authors would like to thank Alison Mitchell for providing the H.E.S.S. images, and Crystal Brogan and Mallory Roberts for providing the 90 cm VLA image. Support for this work by KM was provided by NASA through NuSTAR Cycle
5 Guest Observer Program grant NNH18ZDA001N. SSH acknowledges support by the Natural Sciences and Engineering Research Council of Canada (NSERC) through the Discovery Grants and Canada Research Chairs programs,
and by the Canadian Space Agency (CSA). 
\end{acknowledgments}


\bibliography{citations}{}
\bibliographystyle{aasjournal}



\end{document}